\documentclass{jfm}
\usepackage{graphicx}
\usepackage{epstopdf, epsfig}
\usepackage{amsmath}
\usepackage{color}

\shorttitle{Transport and stratification in the equilibrated Eady model}
\shortauthor{B. Gallet, B. Miquel, G. Hadjerci, K.J. Burns, G. Flierl, R. Ferrari}
\graphicspath{{./figures/}}

\newcommand{\la}{\left<}
\newcommand{\ra}{\right>}
\newcommand{\plh}{%
  {\ooalign{$\phantom{0}$\cr\hidewidth$\scriptstyle\times$\cr}}%
}
\newcommand{\cor}[1]{#1}

\title{Transport and emergent stratification in the equilibrated Eady model: the vortex gas scaling regime}

\author{Basile Gallet\aff{1}
  \corresp{\email{basile.gallet@cea.fr}},
Benjamin Miquel\aff{1,2},
Gabriel Hadjerci\aff{1},
Keaton J. Burns\aff{3},
Glenn Flierl\aff{4},
Raffaele Ferrari\aff{4}}

\affiliation{\aff{1}Universit\'e Paris-Saclay, CNRS, SPEC, CEA, 91191, Gif-sur-Yvette, France.
\aff{2}Univ Lyon, CNRS, Ecole Centrale de Lyon, INSA Lyon, Universit\'e Claude Bernard Lyon 1, LMFA, UMR5509, 69130, Ecully, France.
\aff{3}Department of Mathematics, Massachusetts Institute of Technology, Cambridge, MA 02139, USA.
\aff{4}Department of Earth, Atmospheric, and Planetary Sciences, Massachusetts Institute of Technology, Cambridge, MA 02139, USA.
}

\begin{document}

\maketitle

\begin{abstract}
We numerically and theoretically investigate the Boussinesq Eady model, where a rapidly rotating density-stratified layer of fluid is subject to a meridional temperature gradient in thermal wind balance with a uniform vertically sheared zonal flow. Through a suite of numerical simulations, we show that the transport properties of the resulting turbulent flow are governed by quasi-geostrophic (QG) dynamics in the rapidly rotating strongly stratified regime. The `vortex gas' scaling predictions put forward in the context of the two-layer QG model carry over to this fully 3D system: the functional dependence of the meridional flux on the control parameters is the same, the two adjustable parameters entering the theory taking slightly different values. In line with the QG prediction, the meridional buoyancy flux is depth-independent. The vertical buoyancy flux is such that turbulence transports buoyancy along isopycnals, except in narrow layers near the top and bottom boundaries, the thickness of which decreases as the diffusivities go to zero. The emergent (re)stratification is set by a simple balance between the vertical buoyancy flux and diffusion along the vertical direction. Overall, this study demonstrates how the vortex-gas scaling theory can be adapted to quantitatively predict the magnitude and vertical structure of the meridional and vertical buoyancy fluxes, and of the emergent stratification, without additional fitting parameters.
\end{abstract}

\begin{keywords}
Geophysical and Astrophysical fluid dynamics, rotating stratified flows.
\end{keywords}

\section{Introduction}

The oceans and atmospheres of planets and their satellites are shallow fluid layers subject to the combined effects of rapid global rotation and strong density stratification. Planetary atmospheres are set into motion by differential heating in the meridional \cor{(North-South)} direction -- either as a result of radiative heating by a nearby star or because of heat coming from the deep planetary interior -- while oceans are also subject to mechanical forcing by predominantly zonal \cor{(East-West)} atmospheric winds.

In both cases a meridional temperature gradient emerges and coexists with a vertically sheared zonal flow as a result of thermal wind balance. This base state is unstable, however, and turbulent motion rapidly arises as a consequence of baroclinic instability~\citep{Pedloskybook,Salmonbook,Vallisbook}. The resulting `baroclinic' turbulence enhances heat transport in the meridional direction, which in turn affects the equilibrated meridional temperature profile, but also the local density stratification of the fluid layer. Baroclinic turbulence \cor{strongly contributes to} meridional heat transport at midlatitudes in Earth's atmosphere and other planetary atmospheres, like Jupiter's, and plays a key role in setting the planets' climate~\citep{Liu2010,Read2020}. It is also a dominant feature of ocean currents, most notably in the Southern Ocean, where baroclinic instability of the Antarctic Circumpolar Current (ACC) flowing around Antarctica leads to enhanced meridional transport~\citep{Nowlin1986,Marshall2003,Volkov2010}. The resulting heat and salt transport sets the stratification of the Southern Ocean and to some extent that of all ocean basins~\citep{Wolfe2010,Nikurashin11,Nikurashin12}.

One crucial characteristic of baroclinic turbulence in the Southern Ocean is that its horizontal integral scale is much shorter than the transverse extent of the ACC (a similar scale separation arises in Jupiter's atmosphere but not in the Earth's atmosphere). The scale separation can be leveraged to parameterise the meridional transport in the form of a diffusive closure, where the diffusivity is inferred from a `local' model consisting in an isolated patch of fluid much smaller than the size of the ocean basin but much larger than the integral scale of the flow~\citep{Held99}. A hierarchy of local models can be considered: in the simplest instances the patch of fluid is modelled as a superposition of two vertically invariant layers of fluid stacked in the vertical direction. The equations of motion are further simplified by considering the quasi-geostrophic (QG) approximation to the dynamics of rapidly rotating strongly stratified fluid layers. This leads to the two-layer quasi-geostrophic (2LQG) model, where the governing equations reduce to a conservation equation for the potential vorticity (a scalar field) inside each fluid layer~\citep{Phillips}. Even for this canonical model, however, the transport properties of the equilibrated turbulent flow in the moderate-to-low drag regime have been captured only recently by a scaling theory, which we coined the `vortex gas' scaling regime and extended to the $\beta$-plane~\citep{Gallet2020,Gallet2021}. 

The 2LQG model is a rather crude description of the flow, and one may wonder whether any of these scaling-laws carry over to a fully 3D model of a patch of ocean or atmosphere experiencing weak bottom drag~\citep{Riviere04}. A 3D model also leads to key additional questions that must be addressed to design a skillful parameterisation of turbulent transport: beyond the meridional buoyancy flux, what is the magnitude of the vertical buoyancy flux, i.e., what is the direction of the eddy-induced buoyancy current vector? In a model with a continuous vertical direction, can we predict the vertical structure of the meridional and vertical buoyancy fluxes? Finally, how does this vertical buoyancy flux feed back onto the vertical density stratification?
 
In the present study we thus relax both the finite-layer and the quasi-geostrophic approximations, i.e., we consider a three-dimensional patch of rapidly rotating density-stratified fluid described by the Boussinesq equations. More precisely, we focus on the Boussinesq Eady model, where a uniform meridional temperature gradient coexists with a uniform vertically sheared zonal flow~\citep{Eady49}. We restrict attention to the large-Richardson-number regime \cor{(buoyancy frequency much greater than the vertical shear)} where perturbations amplify through a `geostrophic  instability' mechanism, see~\citet{Stone71}. The departures from the base state thus rapidly evolve into 3D turbulence that we solve using periodic boundary conditions in the horizontal directions. This Boussinesq Eady model is introduced in section~\ref{sec:setup}. We present a suite of numerical simulations of the Eady model in section~\ref{sec:transport}. We show that the numerical results are consistent with QG dynamics. We then recall the prediction of the vortex gas scaling theory for the magnitude of meridional transport and we validate this prediction against the numerical data. In section~\ref{sec:vert}, we extend these predictions to the vertical buoyancy flux and emergent vertical stratification, finding good agreement with the numerical data. In section~\ref{sec:structure}, we turn to the vertical structure of these quantities, which we predict to be depth-invariant within the low-diffusivity QG framework. Based on the Boussinesq data, we show that they are indeed all depth-independent in the bulk of the layer within the low-diffusivity QG regime, but that extremely low diffusivities are needed for that depth-independence to hold near the top and bottom boundaries. The turbulent kinetic energy profile gradually becomes depth-independent as one enters the low-friction vortex-gas regime, in line with a barotropization of the flow. We summarize the results and conclude in section~\ref{sec:conclusion}.
 
\section{The Boussinesq Eady model with bottom drag\label{sec:setup}}

\subsection{Base state}

\begin{figure}
  \centerline{\includegraphics[width=0.8 \textwidth]{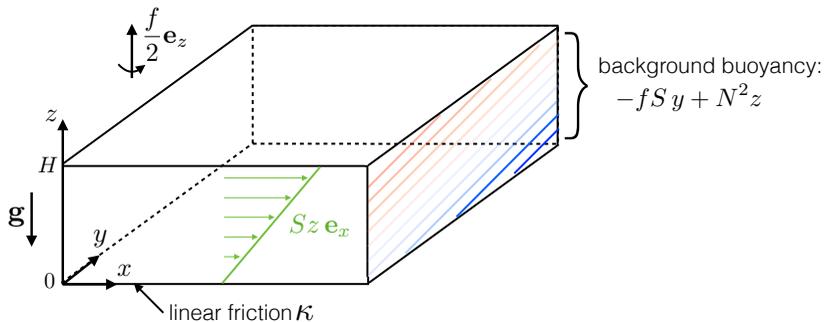}}
  \caption{Base state of the Eady setup: a plane layer of density-stratified fluid is subject to a meridional buoyancy gradient, in a frame rotating at a rate $f/2$ around the vertical axis. A uniformly sheared flow in the zonal direction $x$ coexists with the meridional buoyancy gradient as a result of thermal wind balance. \label{fig:schematic_eady}}
\end{figure}

We consider the Boussinesq equations for a rotating density-stratified layer of fluid inside a 3D domain $(x,y,z)\in[0,L]\times[0,L]\times[0,H]$, with gravity and rotation along the vertical axis $z$:
\begin{eqnarray}
\partial_t {\bf u} + ({\bf u}\cdot {\bnabla}) {\bf u} + f {\bf e}_z \times {\bf u} & = & - \bnabla p  + \alpha g \theta \, {\bf e}_z + {\nu}_\perp \Delta_\perp {\bf u}+ {\nu}_z \partial_{zz} {\bf u}\, , \label{eq:NS}\\
\partial_t \theta + {\bf u}\cdot {\bnabla} \theta & = &  {\nu}_{b;\perp} \Delta_\perp \theta + {\nu}_{b;z} \partial_{zz} \theta  \, , \label{eq:temperature}
\end{eqnarray}
where ${\bf u}(x,y,z,t)$ and $\theta(x,y,z,t)$ denote the velocity and (potential) temperature fields, $p(x,y,z,t)$ is the kinematic pressure, $\alpha$ denotes the thermal expansion coefficient, $g$ denotes gravity. The problem having very different characteristic length scales in the horizontal and vertical directions, we use two different viscosities ${\nu}_z$ and ${\nu}_\perp$ for the vertical and horizontal derivatives arising in the viscous term (i.e., for the $\partial_{zz}$ and $ \Delta_\perp=\partial_{xx}+\partial_{yy}$ terms, respectively). Similarly, we use two different thermal diffusivities for horizontal (${\nu}_{b;\perp}$) and vertical (${\nu}_{b;z}$) thermal diffusion. For simplicity, temperature $\theta$ is considered to be the only stratifying agent in the equations above.

We focus on a steady base state with a vertically sheared zonal flow \cor{${\bf u}=Sz \, {\bf e}_x$} in thermal-wind balance with a temperature gradient along the meridional direction $y$. Substitution into the equations above leads to the following base state:
\begin{eqnarray}
{\bf u}=S z \, {\bf e}_x \, , \qquad \alpha g \, \cor{\theta}(y,z)= - {f S} \,  y + N^2 z\, . \label{basestate}
\end{eqnarray}
The base state is sketched in figure~\ref{fig:schematic_eady}. The meridional temperature gradient in (\ref{basestate}) is directly proportional to the vertical shear $S$ as a result of thermal wind balance, and we have included a \cor{uniform} background vertical temperature gradient \cor{(uniform background buoyancy stratification $N^2$)}. In the following, we thus consider base states both with and without background vertical stratification ($N\neq 0$ and $N=0$, respectively). Strictly speaking, the base state (\ref{basestate}) is a solution to equations (\ref{eq:NS}-\ref{eq:temperature}) only if supplemented with the right boundary conditions. First, we neglect the tiny Ekman layer connecting the shear flow (\ref{basestate}) to the stress-free upper boundary. Secondly, a weak vertical heat flux needs to be imposed at the boundaries to maintain a base state with $N \neq 0$, the magnitude of which vanishes as the vertical diffusivities tend to zero. 
We validate this approach {\it a posteriori} by showing some solutions with $N = 0$. For these solutions a strong density stratification compatible with the top and bottom insulating boundary conditions emerges, and the resulting equilibrated state follows the same scaling behavior as solutions with strong imposed background stratification $N \neq 0$. \cor{In line with the theoretical predictions to come, the emergent stratification is uniform, which justifies the compatibility between simulations performed with and without an imposed uniform background stratification $N^2$.}


\subsection{Departure from the base state: governing equations}

Consider arbitrary departures from the base state:
\begin{eqnarray}
{\bf u}(x,y,z,t) & = & \cor{S z \, {\bf e}_x} + {\bf v} (x,y,z,t) \, , \label{pertv}\\
\alpha g \theta(x,y,z,t) & = & \cor{- {f S} \,  y + N^2 z} +b(x,y,z,t) \, , \label{pertth}
\end{eqnarray}
where $b$ denotes the buoyancy departure. The velocity departure ${\bf v}=(u,v,w)$ is divergence-free in the Boussinesq framework, $\bnabla \cdot {\bf v} = 0$. Substituting (\ref{pertv}-\ref{pertth}) into equations (\ref{eq:NS}-\ref{eq:temperature}) leads to the evolution equations for ${\bf v}$ and $b$:
\begin{eqnarray}
\partial_t {\bf v} + S z \partial_x {\bf v} +  w S {\bf e}_x + ({\bf v} \cdot {\bnabla}) {\bf v} + f {\bf e}_z \times {\bf v} & = & - \bnabla p +  b \, {\bf e}_z + {\nu}_\perp \Delta_\perp {\bf v}+ {\nu}_z \partial_{zz} {\bf v}\, , \label{eq:eqv}\\
\partial_t b - {f S} \, v + N^2 \, w  + S z \partial_x b + {\bf v}\cdot {\bnabla} b & = & {\nu}_{b;\perp} \Delta_\perp b + {\nu}_{b;z} \partial_{zz} b \, . \label{eq:eqtheta}
\end{eqnarray}
We consider periodic boundary conditions for the departure fields ${\bf v}$ and $b$ in the horizontal directions. The top and bottom boundaries are no-flux:
\begin{equation}
\partial_z b|_{z=0} = \partial_z b|_{z=H} = 0 \, ,
\end{equation}
and ${\bf v}$ satisfies free-slip boundary conditions at the top:
\begin{equation}
w|_{z=H} = 0, \qquad \partial_z {\bf v}_\perp|_{z=H} = 0 \, ,
\end{equation}
where ${\bf v}_\perp$ denotes the horizontal components of ${\bf v}$.
Strictly speaking, the bottom boundary condition should be no-slip, ${\bf v}|_{z=0}={\bf 0}$. Such a no-slip boundary condition induces strong friction and pumping associated with the bottom Ekman boundary layer, of typical thickness $\sqrt{\nu_z/f}$. However, the picture of a laminar Ekman layer acting on a perfectly flat ocean bottom is probably a very naive representation of the turbulent boundary layer near the rough ocean floor. We thus turn to a bottom boundary condition that allows us to vary the magnitude of bottom friction independently of the vertical viscosity. We replace the no-slip bottom boundary condition by the following frictional boundary condition:
\begin{eqnarray}
w|_{z=0} = 0 \, , \qquad \partial_z {\bf v}_\perp|_{z=0} = \frac{{\kappa}H}{{\nu}_z} {\bf v}_\perp|_{z=0} \, . \label{bottomBCfirst}
\end{eqnarray}
This boundary condition corresponds to a linear drag law, somewhat similar \cor{to} but weaker than pure Ekman friction. The strength of bottom friction can be tuned through the value of the friction coefficient $\kappa$ (not to be confused with the buoyancy diffusivities ${\nu}_{b;\perp}$ and ${\nu}_{b;z}$). As compared to a standard no-slip boundary condition, (\ref{bottomBCfirst}) allows us to study the regime of weak bottom drag while keeping vertical viscosity values compatible with Direct Numerical Simulation (DNS). 
The boundary condition (\ref{bottomBCfirst}) reduces to a standard no-slip boundary condition for $\kappa \to \infty$, while it reduces to a stress-free boundary condition for $\kappa \to 0$. For arbitrary $\kappa$, the flow near the bottom of the fluid domain resembles a truncated Ekman spiral, with reduced shear and reduced pumping as compared to the standard Ekman boundary layer over a no-slip boundary. In appendix~\ref{app:friction}, we compute the total damping and pumping induced by the boundary condition (\ref{bottomBCfirst}) on the bulk flow.

\subsection{Non-dimensionalization}

We non-dimensionalize the equations using the length scale $H$ and the time scale $f^{-1}$:
\begin{eqnarray}
 & & ({x^\sharp},{y^\sharp},{z^\sharp})=(x,y,z) / H , \qquad ({u^\sharp},{v^\sharp},{w^\sharp}) = (u,v,w)/(fH) , \qquad {t^\sharp} = ft, \\
 & &  {b^\sharp}=b/(f^2 H)  , \qquad {p^\sharp}=p/(f^2 H^2)  , \qquad\, {E}_i=\nu_i/(f H^2) , \qquad {\kappa^\sharp}=\kappa/f \, .
\end{eqnarray}

Dropping the $\sharp$ symbols in the following to alleviate notations, the dimensionless governing equations are:
\begin{eqnarray}
\partial_t {\bf v} + Ro \, z \partial_x {\bf v} +  Ro \, w {\bf e}_x + ({\bf v} \cdot {\bnabla}) {\bf v} + {\bf e}_z \times {\bf v} & = & - \bnabla p + b \, {\bf e}_z + {{E}}_\perp \Delta_\perp {\bf v} + {{E}}_z \partial_{zz} {\bf v} \, , \qquad \label{eq:eqvadim}\\
\partial_t b - Ro \, v + \left(\frac{N}{f}\right)^2 \, w + Ro \, z \partial_x b + {\bf v}\cdot {\bnabla} b & = & {{E}}_{b;\perp} \Delta_\perp b + {{E}}_{b;z} \partial_{zz} b \, , \label{eq:eqbadim}
\end{eqnarray}
where the Rossby number is defined as $Ro=S/f$. The dimensionless boundary conditions are:
\begin{eqnarray}
& & w|_{z=1}=0 \, , \qquad \partial_z {\bf v}_\perp|_{z=1}  =  0 \, , \label{topBC}\\
& & w|_{z=0}=0 \, , \qquad \partial_z {\bf v}_\perp|_{z=0}  =  \frac{{\kappa}}{{{E}}_z} {\bf v}_\perp|_{z=0} \, , \label{bottomBC} \\
& & \partial_z b|_{z=0} = \partial_z b|_{z=1}  =  0 \label{insulatingBC}\, .
\end{eqnarray}

\subsection{Numerical implementation}

We wish to characterize the buoyancy transport achieved by statistically steady solutions to equations (\ref{eq:eqvadim}-\ref{eq:eqbadim}) with the boundary conditions (\ref{topBC}-\ref{insulatingBC}) (see~\citet{Bachman13} for a study of the spin-down problem). Denoting as $\la \cdot \ra$ a time and volume average, we will consider the time- and volume-averaged meridional and vertical buoyancy fluxes, $\la vb \ra$ and $\la wb \ra$, respectively. We also consider the vertical profiles of the meridional and vertical buoyancy fluxes, $\overline{vb}(z)$ and $\overline{wb}(z)$, where the overbar denotes a horizontal area average together with a time average. Finally, a quantity of interest is the emergent stratification $\overline{b}(z)$ that arises in the equilibrated state. The total vertical buoyancy stratification is $(N/f)^2 z+\overline{b}(z)$, and, assuming that the emergent stratification is approximately uniform, we define the Rossby deformation radius $\lambda$ as:
\begin{eqnarray}
\lambda=\sqrt{\left(\frac{N}{f}\right)^2+\overline{b}(1)-\overline{b}(0)} \, ,
\end{eqnarray}
where $\lambda$ is non-dimensionalized with $H$. 

We have performed a suite of numerical simulations of equations (\ref{eq:eqvadim}-\ref{eq:eqbadim}) using Coral, a pseudo-spectral, scalable, time-stepping solver for differential equations~\citep{miquelJOSS}. In Coral, the top and bottom boundary conditions are imposed through basis recombination. The variables are expanded on bases of functions obtained as tensor products of Fourier modes along the horizontal and suitable linear combinations of Chebyshev polynomials, each of which obeys the boundary conditions along $z$. \cor{The divergence-free constraint is readily implemented in Coral by introducing the standard toroidal and poloidal velocity potentials $\psi(x,y,z)$ and $\phi(x,y,z)$, respectively. 
The frictional bottom boundary condition is dealt with by defining modified velocity potentials $\widetilde{\psi}$ and $\widetilde{\phi}$ as $\psi(x,y,z) = P_\psi(z) \widetilde{\psi}(x,y,z)$ and $\phi(x,y,z) = P_\phi(z) \widetilde{\phi}(x,y,z)$, where $P_\psi(z)$ and $P_\phi(z)$ are carefully chosen quadratic polynomials, the coefficients of which depend on the drag coefficient. In particular, these polynomials are tailored so that imposing the Robin boundary conditions~(\ref{bottomBC}) on the velocity field amounts to imposing standard stress-free type boundary conditions for the modified potentials: $\partial_z \widetilde{\psi} =\widetilde{\phi} = \partial_{zz}\widetilde{\phi}=0$. We can thus solve for $\widetilde{\psi}$ and $\widetilde{\phi}$ using standard Galerkin basis recombination.
The code has been benchmarked by careful comparison with results from linear stability analysis, checks that the various power integrals are exactly satisfied in the nonlinear regime, and comparison of the overall buoyancy transport against a run of the same problem performed with the Dedalus solver~\citep{Dedalus}. The numerical runs are initialized with either low-amplitude noise or a checkpoint from a previous simulation. We average the relevant quantities over the statistically steady regime, and make sure that the statistics are correctly converged by comparing with averages performed over the first half of the statistically steady signal only.}


\section{Meridional buoyancy transport \label{sec:transport}}

\subsection{The low-diffusivity regime}

\begin{figure}
  \centerline{\includegraphics[width=0.95 \textwidth]{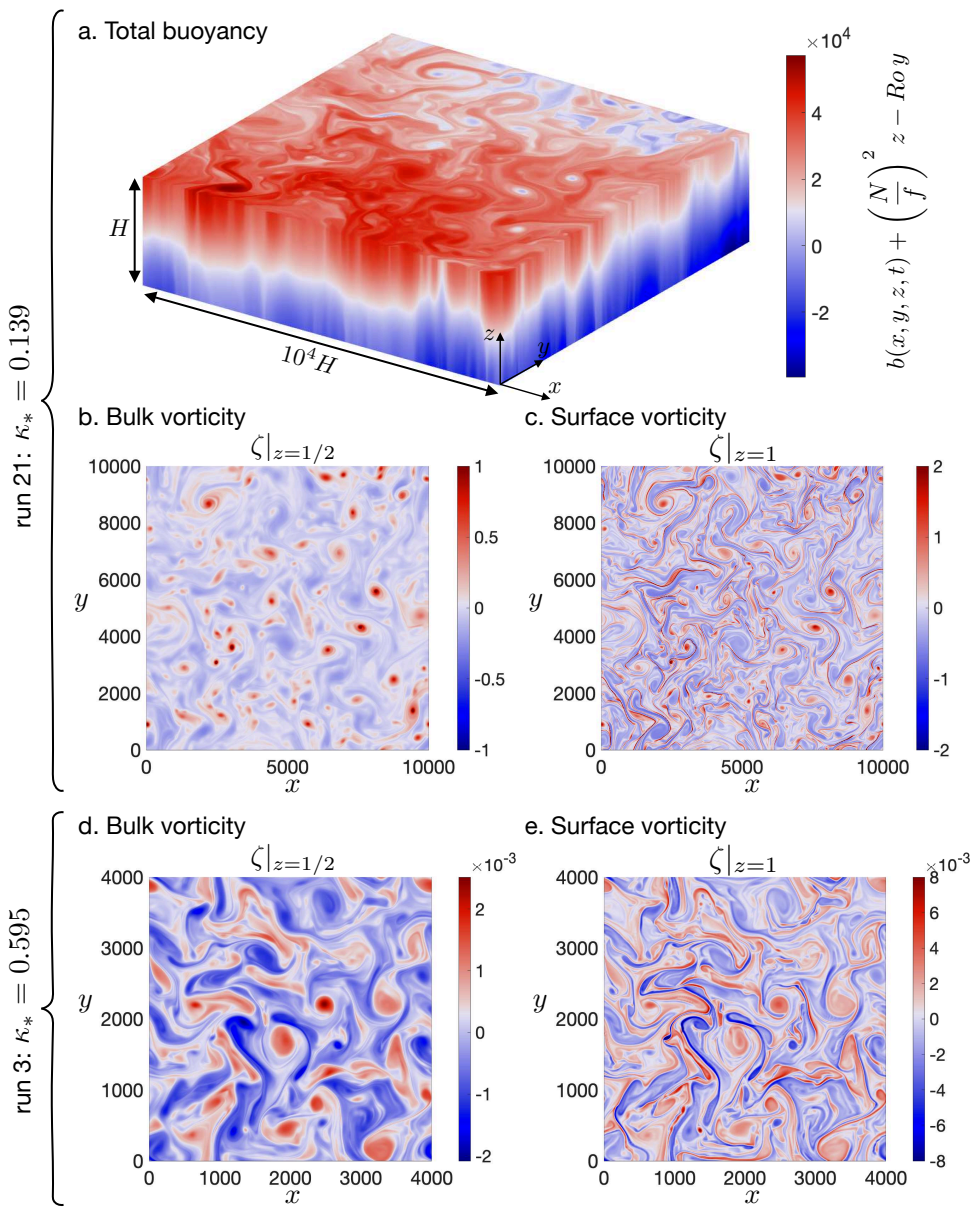}}
  \caption{\cor{Equilibrated baroclinic turbulence illustrated with snapshots from runs 21 (top, low drag) and 3 (bottom, moderate drag). {\bf a:} 3D rendering of the total buoyancy field. Notice the very different vertical and horizontal scales, and thus the very weak isopycnal slopes. {\bf b:} the vertical vorticity at mid-depth shows coherent vortices, with slight cylcone/anticyclone asymmetry in this marginally QG run. {\bf c:} The surface vertical vorticity exhibits sharp frontal structures where relative vorticity greatly exceeds planetary vorticity. {\bf d} and {\bf e:} same as panels b and c, but for larger dimensionless drag $\kappa_*$ in a strongly QG regime. The vorticity at mid-depth shows a less dilute vortex gas than in panel b. QG cyclone/anticyclone symmetry is well satisfied, including in the sharp SQG frontal structures visible in panel e.} \label{fig:snapshots}}
\end{figure}
We focus on the regime where the horizontal extent of the domain is large compared to the energy-containing scale and the mixing-length of the flow \cor{(estimated as the ratio of root-mean-square horizontal buoyancy fluctuations divided by the meridional buoyancy gradient, see e.g. \citet{Thompson06})}. \cor{Through variations of the domain size, we checked that} the transport properties of the flow are then independent of the horizontal domain size \cor{(for brevity, we do not report these validation runs in the Tables below)}. The diffusivities are chosen to be equal for momentum and buoyancy, i.e., ${{E}}_{b;\perp}={{E}}_{\perp}$ and ${{E}}_{b;z}={{E}}_{z}$ in all of the numerical runs. This corresponds to setting the Prandtl number to unity, an appropriate choice in the present context where the viscosities and diffusivities represent mixing by small-scale turbulence, and not molecular processes. The horizontal diffusivities are small enough not to affect the transport properties of the equilibrated flow. The role of the vertical diffusivities is more subtle: anticipating the results in sections~\ref{sec:transport} and~\ref{sec:vert}, we choose small enough vertical diffusivities for the transport properties of the flow to be well-described by bulk diffusivity-free QG dynamics with bottom friction. In that regime, however, the emergent stratification depends on the vertical buoyancy diffusivity, as described in section~\ref{sec:emergent}. Table~\ref{Tableruns} lists the various control parameters for the numerical runs considered in the present study: we vary the background stratification, the Rossby number, the diffusivities and the friction coefficient. 

The qualitative aspect of the equilibrated flow is illustrated in Figure~\ref{fig:snapshots}, where we show a 3D rendering of the total buoyancy field together with horizontal slices of the dimensionless vertical vorticity $\zeta$ at mid-depth and at the top surface. We observe that the flow remains strongly stratified in the equilibrated state, with weak isopycnal slopes in the bulk of the fluid domain. \cor{Panels a, b and c correspond to a run that is marginally QG and has very low dimensionless drag. The bulk vertical vorticity exhibits a relatively dilute gas of vortices, although there is some asymmetry between cyclones and anticyclones. This asymmetry probably stems from the sharp non-QG frontal structures that develop at the surface~\citep{Hoskins72,Klein08,Ragone16,Siegelman2020}, see panel c.  We will see in the following that the transport properties of this flow are correctly described by QG theory despite the slight cyclone/anticyclone asymmetry of the interior flow. Panels d and e illustrate a run in the strongly QG regime with moderately low drag. The vortex gas is less dilute as a result of the larger drag. QG cyclone-anticyclone symmetry is well-satisfied both at mid-depth (panel d) and in the surface quasi-geostrophic (SQG) fronts visible in panel e.}


\begin{figure}
  \centerline{\includegraphics[width=0.65 \textwidth]{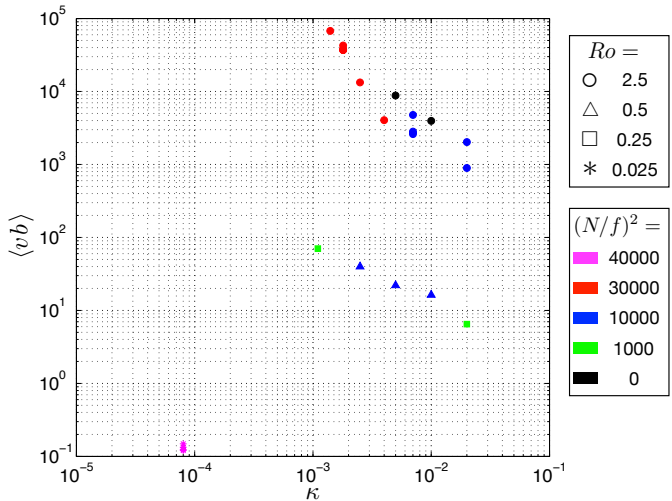}}
  \caption{Raw meridional buoyancy flux as a function of the friction coefficient for the 22 numerical runs. The symbols corresponds to different values of the Rossby number, while colour codes for the background stratification. The meridional flux increases with decreasing friction coefficient for otherwise constant parameters.\label{fig:raw_data}}
\end{figure}

In Figure~\ref{fig:raw_data}, we show the overall meridional buoyancy flux $\la vb \ra$ as a function of the friction coefficient $\kappa$. The buoyancy flux varies over more than four orders of magnitude over the entire data set. Meridional transport increases as the friction coefficient decreases (for otherwise constant parameters), and it decreases rapidly with decreasing Rossby number.

\subsection{Bulk QG dynamics: further reducing the number of control parameters}

The qualitative features of the equilibrated flows in Figure~\ref{fig:snapshots} point to QG dynamics. One can derive the QG approximation to the Boussinesq Eady system following the standard procedure reviewed in appendix~\ref{app:QG} (see also~\citet{Salmonbook,Vallisbook}). Decompose the pressure field into a time and horizontal mean $\overline{p}(z)$ plus fluctuations $\tilde{p}(x,y,z,t)$: 
\begin{eqnarray}
{p}(x,y,z,t)=\overline{p}(z)+\tilde{p}(x,y,z,t), \qquad \text{with} \quad \overline{\tilde{p}}=0 \, . \label{horizontalmeandecomp}
\end{eqnarray}
Notice that the horizontal and time derivatives of $p$ and $\tilde{p}$ are equal, so that $p$ and $\tilde{p}$ can be used interchangeably in many (but not all) of the expressions to come. 
Assuming that the emergent stratification is approximately uniform and neglecting the diffusive terms in the bulk of the flow, the potential vorticity (PV) conservation equation reads (see appendix~\ref{app:QG}): 
\begin{equation}
\partial_t q + J(\tilde{p},q) + Ro \, z \, \partial_x q  =  0 \, , \label{PVbulk}
\end{equation}
where the PV is:
\begin{equation}
q = \bnabla_\perp^2 \tilde{p} +  \frac{\partial_{zz} \tilde{p}}{\lambda^2} \, . \label{QGPV}
\end{equation}
The boundary conditions are the buoyancy equation written at the boundaries, remembering the hydrostatic relation ${b}=\partial_z {p}$. 
At the top free-surface one obtains:
\begin{eqnarray}
\partial_{tz} \tilde{p}|_1 - Ro \, \partial_x \tilde{p}|_1 + Ro \, \partial_{xz} \tilde{p}|_1 + J(\tilde{p}|_1,\partial_z \tilde{p}|_1) & = & 0 \label{BCQG1} \, .
\end{eqnarray}
The second boundary condition stems from the buoyancy equation written just above the partial Ekman layer connecting the bulk flow to the frictional bottom boundary condition (a height denoted as $z=0^+$):
\begin{eqnarray}
\partial_{tz} \tilde{p}|_{0^+} - Ro \, \partial_x \tilde{p}|_{0^+} + J(\tilde{p}|_{0^+},\partial_z \tilde{p}|_{0^+}) + \lambda^2 w|_{0^+}& = & 0  \, .
\end{eqnarray}
where $w|_{0^+}$ denotes the vertical pumping velocity induced by the partial Ekman layer. As shown in appendix~\ref{app:friction}, this pumping velocity takes the form:
\begin{eqnarray}
w|_{0^+} = \kappa_\text{eff} \zeta|_{0^+} =\kappa_\text{eff} \Delta_\perp \tilde{p}|_{0^+} \, , \label{pumpingBC}
\end{eqnarray}
where the effective friction coefficient acting on the interior flow is:
\begin{eqnarray}
\kappa_\text{eff} = \frac{\sqrt{2 {E}_z}}{1+ \frac{\sqrt{2 {E}_z}}{\kappa}+\frac{{E}_z}{\kappa^2}} \times \left(\frac{1}{2}+ \sqrt{\frac{{E}_z}{2}} \frac{1}{\kappa} \right) \, , \label{eq:defkappaeff}
\end{eqnarray}
both $\kappa$ and $\kappa_\text{eff}$ being non-dimensionalized with $f$. \cor{Expression (\ref{eq:defkappaeff}) reduces to the standard Ekman friction coefficient in the no-slip limit $\kappa \to \infty$ \citep{Pedloskybook}.} The buoyancy equation at $z=0^+$ finally reads:
\begin{eqnarray}
\partial_{tz} \tilde{p}|_{0^+} - Ro \, \partial_x \tilde{p}|_{0^+} + J(\tilde{p}|_{0^+},\partial_z \tilde{p}|_{0^+}) + \lambda^2 \kappa_\text{eff} \Delta_\perp \tilde{p}|_{0^+}& = & 0 \label{BCQG0} \, .
\end{eqnarray}
QG dynamics is governed by the conservation equation (\ref{PVbulk}) for the bulk PV (\ref{QGPV}), with the top and bottom boundary conditions (\ref{BCQG1}) and (\ref{BCQG0}). The fluid communicates in the vertical direction through slight vertical displacements of the isopycnals, which induce modifications in vertical vorticity $\zeta=\Delta_\perp p$ through the conservation of PV (for the specific case of a cylindrical fluid column contained between two isopycnals, this process can be understood as the conservation of angular momentum as the column gets compressed or stretched). Diffusion plays no roles in this process, the consequence being that $\kappa$ and ${E}_z$ enter the QG dynamics only through the effective friction coefficient $\kappa_\text{eff}$ associated with pumping in the bottom (partial) Ekman layer.
The number of dimensionless parameters relevant to the QG regime can be further reduced through the introduction of the QG scalings:
\begin{eqnarray}
t= \frac{\lambda}{Ro} T \, , \quad x= \lambda X, \quad y= \lambda Y \, , \quad \tilde{p}=Ro \lambda \, P(X,Y,z,T) \, , \quad q = \frac{Ro}{\lambda} Q(X,Y,z,T) \, .   
\end{eqnarray}
Through this change of variables the QG equations become:
\begin{eqnarray}
&  & \partial_T Q + J_{\bf X}(P,Q)+z \partial_X Q  = 0 \, , \label{eq:rescaledQG1}\\
& & Q  =  \Delta_{{\bf X},\perp} P + \partial_{zz}P \, , \label{eq:rescaledQG2}\\
& & \partial_{Tz} P|_1 - \partial_X P|_1 +  \partial_{Xz} P|_1 + J_{\bf X}(P|_1,\partial_z P|_1) =  0 \, , \label{eq:rescaledQG3}\\
& & \partial_{Tz} P|_{0^+} - \partial_X P|_{0^+} + J_{\bf X}(P|_{0^+},\partial_z P|_{0^+}) + \frac{\kappa_\text{eff} \lambda}{Ro} \Delta_{{\bf X},\perp} P|_{0^+} =  0 \, , \label{eq:rescaledQG4}
\end{eqnarray}
where $J_{\bf X}(f,g)=\partial_X f \partial_Y g-\partial_Y f \partial_X g$ and $\Delta_{{\bf X},\perp} = \partial_{XX}+ \partial_{YY}$.
An instructive feature of this set of equations is that it involves a single control parameter:
\begin{eqnarray}
\kappa_*= \frac{\kappa_\text{eff} \lambda}{Ro} \, ,
\end{eqnarray}
which is the Boussinesq equivalent of the relevant dimensionless drag coefficient in the QG framework~\citep{Thompson06,Gallet2020}. The equilibrated solutions to the set of equations (\ref{eq:rescaledQG1}-\ref{eq:rescaledQG4}) are characterized by a meridional buoyancy transport $D_*=\la \partial_X P \partial_z P \ra=\la v b \ra/(Ro^2 \lambda)$ that depends only on $\kappa_*$. We conclude that:
\begin{eqnarray}
D_* \equiv \frac{\la v b \ra}{Ro^2 \lambda}= {\cal F} \left( \kappa_* = \frac{\kappa_\text{eff} \lambda}{Ro} \right) \, , \label{eq:QGrescaling}
\end{eqnarray}
where the unknown function ${\cal F}$ denotes the dependence of the QG buoyancy eddy diffusivity $D_*$ on the QG dimensionless drag $\kappa_*$. As discussed in \citet{Thompson06} and \citet{Gallet2020}, the 2LQG equivalent of $D_*$ readily quantifies the ratio of the meridional buoyancy flux over the meridional buoyancy gradient, hence the name `diffusivity'. Indeed, while the change of variables above provides a rigorous way to identify $D_*$ and $\kappa_*$ as the `order' and `control' parameters of the QG regime, the form of $D_*$ and $\kappa_*$ could have been guessed from studies of the strongly idealized 2LQG model: as shown in~\citet{Thompson06}, the 2LQG model on the $f$-plane reduces to the study of $D_*$ versus $\kappa_*$, where $D_*$ is the meridional flux divided by the product of the deformation radius with the squared shearing velocity, while $\kappa_*$ is the bottom drag coefficient multiplied by the deformation radius and divided by the shearing velocity.

\begin{figure}
  \centerline{\includegraphics[width=0.7 \textwidth]{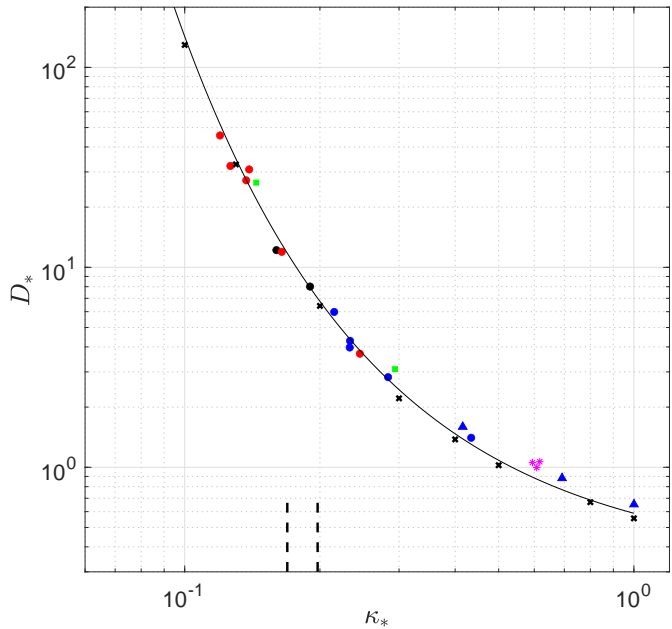}}
  \caption{Dimensionless eddy-induced diffusivity as a function of the dimensionless effective friction coefficient, using the QG non-dimensionalization. The entire data set falls onto a master curve, indicating that QG dynamics governs the overall buoyancy transport. Same symbols and colours as in Figure~\ref{fig:raw_data}. The thick vertical dashed lines indicate the predicted values of $\kappa_*$ according to equation (\ref{eq:predictedstrat}) for the runs without background stratification (black circles). The predicted values of $\kappa_*$ agree with the DNS ones within $5\%$. Black crosses are the solutions to the QG system. The solid line is the vortex-gas prediction (\ref{eq:diffusivityVG}) with $c_1=0.32$ and $c_2=0.61$. \label{fig:Dstar_vs_kappastar}}
\end{figure}

Equation (\ref{eq:QGrescaling}) suggests a quantitative way to assess whether meridional buoyancy transport is governed by QG dynamics in the present simulations of the Boussinesq system: in Figure~\ref{fig:Dstar_vs_kappastar}, we plot $D_*$ as a function of $\kappa_*$ for the entire dataset. This representation leads to a collapse onto a master curve of the entire data set, which includes various values of the Rossby number, background stratification and vertical Ekman numbers. This confirms the validity of the QG scalings for bulk transport in the present numerical data, despite the emergence of potentially non-QG frontal dynamics near the boundaries. 

We further compare the results of the Boussinesq system to QG theory by solving the QG Eady problem explicitly. A trivial solution to the QG PV conservation equation (\ref{eq:rescaledQG1}) is $Q=0$. In appendix~\ref{app:QG}, we recall how this constraint leads to an inversion relation providing $P|_{1}$ and $P|_{0^+}$ in terms of $\partial_z P|_{1}$ and $\partial_z P|_{0^+}$ at each time step. Solving the 3D QG problem then reduces to solving the coupled equations (\ref{eq:rescaledQG3}) and (\ref{eq:rescaledQG4}) for the 2D fields $\partial_z P|_{1}$ and $\partial_z P|_{0^+}$, using the inversion relation to infer $P|_{1}$ and $P|_{0^+}$ at each time step. \cor{The approach is very similar to the surface quasi-geostrophy (SQG)~\citep{Blumen78,Held95,Callies16,Lapeyre17}, the difference being that there are two horizontal boundaries in the present system (at top and bottom), as opposed to a single horizontal boundary in the standard SQG system. The second boundary crucially allows for the emergence of depth-invariant barotropic eddies in the present system.}
The effectively 2D equations are solved on a GPU in a large-enough domain for finite-size effects to be negligible. To damp the small-scale filaments, we also add hyperviscous terms $-\mu \Delta_{{\bf X},\perp}^4  \partial_{z} P|_1$ and $-\mu \Delta_{{\bf X},\perp}^4  \partial_{z} P|_{0^+}$ to the right-hand sides of (\ref{eq:rescaledQG3}) and (\ref{eq:rescaledQG4}), respectively, where the dimensionless hyperdiffusivity $\mu$ is chosen small enough to not affect the meridional heat flux in the equilibrated state. The resulting data points are plotted using crosses in Figure~\ref{fig:Dstar_vs_kappastar}. They agree very well with the 3D Boussinesq data, \cor{indicating that PV vanishes in the interior of the 3D domain as a result of the dissipative terms, with negligible PV injection by frontal structures at the boundaries (in contrast with zonally invariant 2D primitive-equation models, see~\citet{Nakamura89} and \citet{Garner92})}.


To summarize, we have established that meridional transport in the equilibrated Eady Boussinesq system is governed by QG dynamics, as illustrated by the collapse of the Boussinesq data in the representation of Figure~\ref{fig:Dstar_vs_kappastar} and their good agreement with solutions of the QG system. Developing a theory for the magnitude of the meridional transport in the equilibrated Eady problem then reduces to the determination of the function ${\cal F}$ in the scaling relation (\ref{eq:QGrescaling}).

\subsection{Vortex gas scaling theory}

We recently introduced such a scaling theory, coined the vortex gas scaling regime~\citep{Gallet2020}, within the framework of the 2LQG model. The central idea is that the barotropic flow consists in a dilute gas of isolated vortices wandering around through mutual advection~\citep{Thompson06,Carnevale}. The vortex cores are small compared to the inter-vortex distance, with a radius comparable to the Rossby deformation radius. Buoyancy fluctuations arise through the distortion of the background meridional gradient by the vortical flow. Focussing on a single barotropic vortex dipole shows that the mixing length is comparable to the inter-vortex distance, while the diffusivity is given by the product of the inter-vortex distance with the velocity at which the vortex cores move. Two energetic arguments complete the scaling theory: first, a typical velocity scale is estimated by equating the potential energy drop with the final kinetic energy as a fluid column travels over one mixing-length (seen as the mean-free path of the turbulent fluid motion). This `slantwise free-fall' argument yields the typical velocity of a fluid column located between two vortices, which is also the velocity scale at which the vortex cores move. Second, at equilibrium the energy released through baroclinic instability must be balanced by frictional dissipation. Significantly larger velocities arise at the immediate periphery of the vortex cores, where fluid particles circle rapidly without transporting much buoyancy. These large velocities are associated with large frictional dissipation, with important consequences for the energy power integral, which is the last scaling relation of the theory. This line of arguments provides a theoretical expression for the function ${\cal F}$ in (\ref{eq:QGrescaling}). Specifically, for linear bottom friction we obtain:
\begin{eqnarray}
D_* & = & c_1 \exp \left({\frac{c_2}{\kappa_*}}\right) \, , \label{eq:diffusivityVG}
\end{eqnarray}
where $c_1$ and $c_2$ are adjustable parameters that may depend on details of the specific model. Based on low-drag simulations of the 2LQG model with equal layer depths, we estimated $c_1^{(2LQG)}\simeq 2.0$ and $c_2^{(2LQG)}\simeq 0.72$~\citep{Gallet2021}. For the present Boussinesq Eady data, we show in Figure~\ref{fig:Dstar_vs_kappastar} that the vortex gas prediction (\ref{eq:diffusivityVG}) accurately captures the master curve, with $c_1=0.32$ and $c_2=0.61$. This indicates that the physical intuition gathered from the 2LQG model carries over to the fully 3D Eady model.

\section{Vertical transport and emergent stratification\label{sec:vert}}

\begin{figure}
\centerline{\includegraphics[width=0.7 \textwidth]{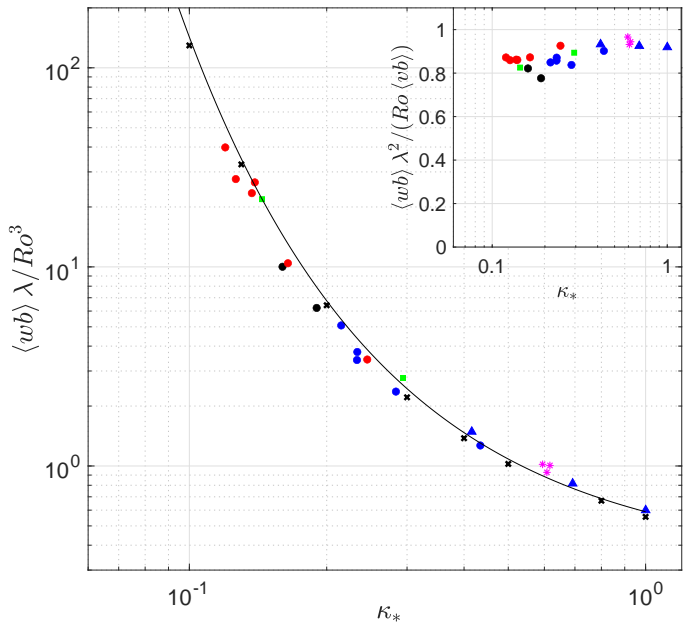}}
  \caption{{\bf Inset:} Ratio ${r}$ between the two sides of the approximate relation (\ref{eq:GMrelation}). This ratio is always close to unity, with a typical value ${r}\simeq 0.85$ over the entire data set. This ratio approaches unity as the diffusivities are lowered, indicating along-isopycnal transport. {\bf Main figure:} Dimensionless vertical buoyancy flux as a function of the dimensionless effective friction coefficient, both being non-dimensionalized following the QG scalings. Combining the vortex-gas theory with pure along-isopycnal transport leads to the prediction on the right-hand side of (\ref{eq:QGwb}), shown as a solid line. Same symbols and colours as in Figure~\ref{fig:raw_data}.  \label{fig:wb_vs_kappastar}}
\end{figure}

As compared to the 2LQG system, the Eady model includes a continuous vertical direction. The eddy-induced buoyancy current is a vector with both a meridional and a vertical component. The vertical buoyancy flux is crucial because it plays a central role in setting the emergent vertical stratification. In the following we propose simple scaling arguments to quantitatively predict the vertical buoyancy flux before computing the emergent stratification. We will show that the latter can be negligible, comparable, or much greater than the background stratification depending on the parameter regime. In particular, we will show that restratification can be strong enough to induce QG dynamics even in the absence of an imposed background vertical stratification.

\subsection{Bulk eddy transport is along isopycnals}

In a similar fashion to (\ref{horizontalmeandecomp}), decompose the buoyancy field into a time and horizontal mean $\overline{b}(z)$ plus fluctuations $\tilde{b}(x,y,z,t)$: 
\begin{eqnarray}
{b}(x,y,z,t)=\overline{b}(z)+\tilde{b}(x,y,z,t), \qquad \text{with} \quad \overline{\tilde{b}}=0 \, .
\end{eqnarray}
Upon multiplying the buoyancy equation (\ref{eq:eqbadim}) with $b$ before averaging temporally and horizontally, assuming that the fluctuations are much weaker than the mean, $\tilde{b} \ll \overline{b}$, one obtains:
\begin{eqnarray}
-Ro \, \overline{vb} + \left[ \left( \frac{N}{f}\right)^2 + \partial_z \overline{b} \right] \overline{wb} \simeq -{{E}}_{b;\perp} \overline{|\bnabla_{\perp} b|^2} + {{E}}_{b;z} \overline{b \, \partial_{zz}b} \, . \label{eq:horizavgb}
\end{eqnarray}
If the diffusive terms on the right-hand side are negligible, we recover the standard conclusion that the eddy-induced buoyancy current is along isopycnals~\citep{Gent90,Young12}. In terms of the overall magnitude of the vertical buoyancy flux, this leads to the simple relation:
\begin{eqnarray}
\la wb \ra \simeq \frac{Ro}{\lambda^2} \, \la vb \ra \, , \label{eq:GMrelation}
\end{eqnarray}
the slope of the isopycnals being $Ro/\lambda^2$. In Figure~\ref{fig:wb_vs_kappastar} we plot the ratio of the left-hand side over the right-hand side of equation (\ref{eq:GMrelation}), denoted as ${r}$, for solutions of the Boussinesq system. This ratio is indeed approximately constant at low drag, with a value ${r} \simeq 0.85$ close to unity, the slight departure from one being due to the small diffusive contributions in (\ref{eq:horizavgb}). The ratio ${r}$ approaches unity if the \cor{diffusive} terms are small enough in a regime where the assumptions of QG are well satisfied. \cor{Solutions to the QG system -- equations (\ref{eq:rescaledQG3}-\ref{eq:rescaledQG4}) together with inversion relations (\ref{inversion1}-\ref{inversion2}) -- satisfy ${r}=1$ exactly.} Overall, the combination of the vortex-gas scaling prediction with the Gent-McWilliams argument leads to a good theoretical prediction for the vertical buoyancy flux:
\begin{eqnarray}
\la wb \ra = c_1 {r} \, \frac{Ro^3}{\lambda} \exp \left({\frac{c_2}{\kappa_*}}\right) \simeq c_1 \, \frac{Ro^3}{\lambda} \exp \left({\frac{c_2}{\kappa_*}}\right) \, , \label{eq:QGwb}
\end{eqnarray}
as illustrated in Figure~\ref{fig:wb_vs_kappastar}. 

\subsection{Strength of the emergent stratification\label{sec:emergent}}

One then estimates the emergent stratification from the vertical buoyancy flux through the potential energy evolution equation. Multiplying the buoyancy equation (\ref{eq:eqbadim}) by $z$ before averaging over space and time yields, after \cor{using $\overline{w}=0$ and performing} some integrations by parts using the boundary conditions:
\begin{eqnarray}
Ro \int_0^1 z \overline{v} \, \mathrm{d}z + \la wb \ra & = & {E}_{b; z} \left[ \overline{b}(1)-\overline{b}(0) \right] \, . \label{eq:PEequation}
\end{eqnarray}
We wish to show that the term involving the mean meridional flow $\overline{v}(z)$ is negligible as compared to the vertical buoyancy flux on the left-hand side. This mean meridional velocity corresponds to an ageostrophic flow, because a mean zonal buoyancy gradient maintaining $\overline{v}(z)$ through thermal wind balance would be incompatible with the periodic boundary conditions. The profile $\overline{v}(z)$ can be estimated by time- and area-averaging the zonal component of the momentum equation (\ref{eq:eqv}):
\begin{eqnarray}
\overline{v} = \partial_z \overline{wu} - E_z \partial_{zz} \overline{u} \, . \label{estimatev}
\end{eqnarray}

The second term on the right-hand side is negligible in the bulk of the domain because $E_z \ll 1$. The first term on the right-hand side involves $\overline{wu}$. It leads to a contribution $-Ro \la wu \ra$ to the left-hand side of (\ref{eq:PEequation}), to be compared with the second term, $\la wb \ra$. A crude estimate of the ratio of these two contributions is $Ro { \la u^2 \ra^{1/2}}/{\la b^2 \ra^{1/2}}$, which using thermal-wind balance for a flow of horizontal scale $\lambda$ is of order $Ro /\lambda \ll 1$. This simple estimate is a first indication that the $\overline{v}$ term in equation (\ref{eq:PEequation}) is negligible as compared to the vertical buoyancy flux.

We stress the fact, however, that $\la wu \ra$ and $\overline{v}$ are even smaller than the crude estimate above. Indeed, to lowest order the magnitude of $\la wu \ra$ can be estimated using QG theory, the diagnostic vertical velocity being inferred from equation (\ref{eq:appborder1}). One can then leverage the cyclone-anticyclone symmetry of the QG system to show that the resulting $\overline{wu}$ vanishes, as explained in appendix~\ref{app:sym}. In other words, while QG dynamics induces a nonzero vertical buoyancy flux $\overline{wb}$ in the bulk of the domain, it leads to a vanishing vertical flux of zonal momentum $\overline{wu}$ for symmetry reasons. Nonzero $\overline{wu}$ in the bulk may only arise at next order in the QG expansion, where cyclone-anticyclone symmetry is broken~\citep{Muraki99,Hakim02,Gallet14}. We conclude that $\la wu \ra$ is smaller than the estimate of the previous paragraph by (at least) a factor corresponding to the isopycnal slope, $Ro/\lambda^2 \ll 1$. The $\overline{v}$ term is thus fully negligible as compared to $\la wb \ra$ in equation~(\ref{eq:PEequation}).

Having shown that the dominant balance in equation (\ref{eq:PEequation}) is between the vertical buoyancy flux and the diffusive flux associated with the emergent stratification, we obtain the following expression for the emergent stratification:
\begin{eqnarray}
\overline{b}(1)-\overline{b}(0) \simeq \frac{\la wb \ra}{{E}_{b; z}} \, . \label{eq:emergentstrat}
\end{eqnarray}
One should notice that the vertical buoyancy diffusivity comes back into play through the coefficient ${E}_{b; z}$ \cor{in} the denominator. While the vertical momentum diffusivity (the viscosity coefficient) affects the scaling behavior of the system only marginally, through a modification of the effective friction coefficient (\ref{eq:defkappaeff}), the vertical buoyancy diffusivity greatly impacts the emergent stratification (\ref{eq:emergentstrat}). The transport properties of the system are independent of the vertical buoyancy diffusivity only in the regime where the emergent stratification (\ref{eq:emergentstrat}) is negligible as compared to the imposed background stratification. Otherwise, changes in ${E}_{b; z}$ induce changes in the emergent stratification (\ref{eq:emergentstrat}) and thus in the Rossby deformation radius $\lambda$, the consequence being that the transport properties of the equilibrated state are shifted following the QG master curves in Figures~\ref{fig:Dstar_vs_kappastar} and \ref{fig:wb_vs_kappastar}. From equation (\ref{eq:emergentstrat}), we can estimate the emergent stratification directly in terms of the control parameters of the problem. While this can be done for an arbitrary background stratification, we focus on two limiting situations: the case without imposed background stratification, and the case where the emergent stratification is negligible as compared to the imposed background one.

In the absence of imposed background stratification, \cor{where $\overline{b}(1)-\overline{b}(0)=\lambda^2$, }we insert expression (\ref{eq:QGwb}) for the vertical flux into (\ref{eq:emergentstrat}), before writing the resulting relation as:
\begin{eqnarray}
  \left(\frac{c_2}{\kappa_*} \right)^3 e^{{c_2}/{\kappa_*}} = \frac{c_2^3}{c_1} \times \frac{{E}_{b; z} }{\kappa_\text{eff}^3} \, .
\end{eqnarray}
We invert this relation to obtain ${c_2}/{\kappa_*}$, which leads to the following expression for the emergent Rossby deformation radius (emergent stratification):
\begin{eqnarray}
\lambda = \frac{c_2 Ro}{3 \kappa_\text{eff} {\cal W}\left[ \frac{c_2 }{3 c_1^{1/3}} \times  \frac{ {E}_{b; z}^{1/3}}{\kappa_\text{eff}} \right]}\, , \text{that is}\quad  \kappa_* = \frac{c_2}{3  {\cal W}\left[ \frac{c_2 }{3 c_1^{1/3}} \times  \frac{ {E}_{b; z}^{1/3}}{\kappa_\text{eff}} \right]} \, , \label{eq:predictedstrat}
\end{eqnarray}
where ${\cal W}$ denotes the Lambert function. These equations relate emergent quantities (on the left-hand side) to control parameters only (on the right-hand side). The first expression in (\ref{eq:predictedstrat}) provides the emergent stratification, while the second one readily gives the dimensionless friction $\kappa_*$ arising in the QG framework (the abscissa in Figures~\ref{fig:Dstar_vs_kappastar} and \ref{fig:wb_vs_kappastar}). We have performed two numerical runs without background stratification to validate these predictions. In Figure~\ref{fig:Dstar_vs_kappastar}, we show that the resulting meridional flux falls onto the master curve, at an abscissa given by the theoretical expression (\ref{eq:predictedstrat}) within $5\%$ accuracy.

\begin{figure}
\centerline{\includegraphics[width= 0.6 \textwidth]{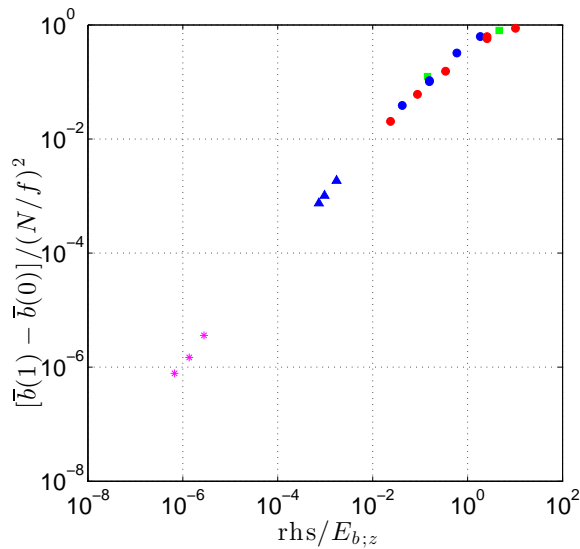}}
  \caption{Ratio of the emergent stratification, $\overline{b}(1)-\overline{b}(0)$, over the background one, $(N/f)^2$, versus the ratio of the right-hand side of (\ref{criterionstrat}) over $E_{b;z}$. In agreement with the criterion (\ref{criterionstrat}), the emergent stratification is negligible as compared to the background one when the abscissa is much less than unity. In this regime the abscissa readily provides the theoretical prediction for the emergent stratification divided by the background one, and indeed the data points fall onto the diagonal. The emergent stratification becomes comparable to the background one as the abscissa approaches unity. Same symbols and colours as in Figure~\ref{fig:raw_data}.  \label{fig:emergentstrat}}
\end{figure}

We now consider the case of a nonzero imposed background stratification $(N/f)^2$. We wish to determine when the emergent stratification is negligible as compared to the imposed one. To wit, we estimate the emergent stratification by inserting into (\ref{eq:emergentstrat}) expression (\ref{eq:QGwb}) for the vertical flux, in which we replace $\lambda$ with $N/f$. Demanding that this emergent stratification be negligible as compared to the imposed one, $(N/f)^2$, then amounts to satisfying the inequality:
\begin{eqnarray}
\quad {E}_{b; z}  \gg c_1 \frac{Ro^3}{(N/f)^3} \exp \left( \frac{c_2 Ro}{\kappa_\text{eff} N/f} \right) \, . \label{criterionstrat}
\end{eqnarray}
The criterion (\ref{criterionstrat}) is expressed in terms of the control parameters of the system only. It allows one to assess {\it a priori} whether or not the emergent stratification is negligible as compared to the background one, as illustrated in Figure~\ref{fig:emergentstrat}. When the inequality (\ref{criterionstrat}) is satisfied, the emergent stratification is negligible and the meridional and vertical buoyancy fluxes are readily obtained by replacing $\lambda$ with $(N/f)$ in expressions (\ref{eq:diffusivityVG}) and (\ref{eq:QGwb}).

To summarize, combining the vortex-gas scaling theory with both the along-isopycnal-transport argument and a simple advective-diffusive buoyancy-flux balance along the vertical direction, we obtain theoretical expressions for the meridional buoyancy flux, the vertical buoyancy flux and the emergent stratification in terms of the control parameters of the problem. 

\section{Vertical structure\label{sec:structure}}

\begin{figure}
\centerline{\includegraphics[width= \textwidth]{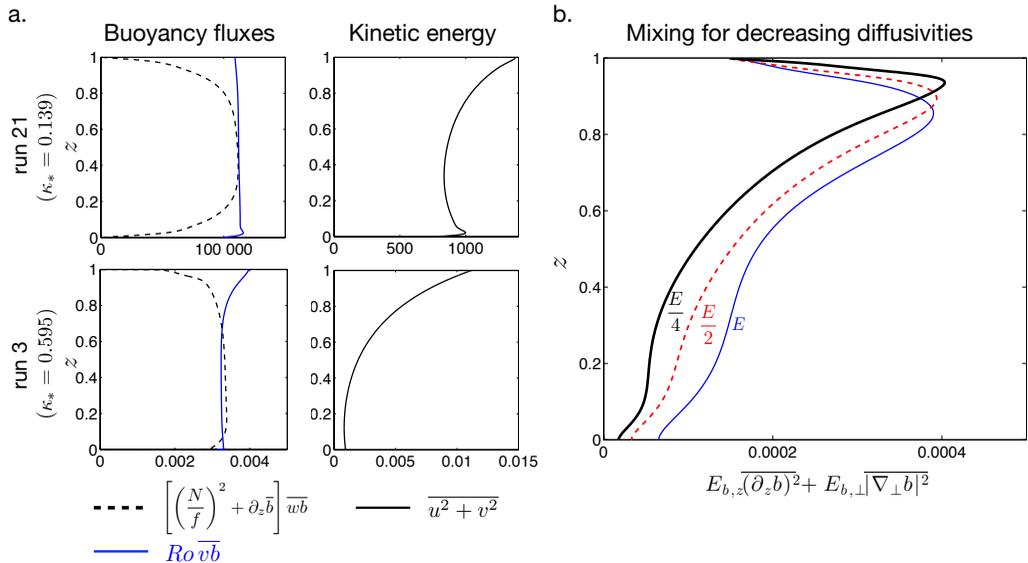}}
  \caption{{\bf a.} Profiles of the scaled meridional and vertical buoyancy fluxes arising in equation (\ref{eq:horizavgb}), and profiles of horizontal turbulent kinetic energy, for runs 21 (top panels, $\kappa_*=0.139$) and 3 (bottom panels, $\kappa_*=0.595$). For both runs the meridional flux is approximately depth-independent, in agreement with the QG prediction. The top panels correspond to the low-drag regime, with moderately low Ekman numbers. The vertical flux retains some depth-dependence near the boundaries. By contrast, for the very-low-diffusivity run of the bottom panels the vertical buoyancy flux is approximately depth-invariant. The TKE profiles show the gradual barotropization of the flow as we go from moderately low drag (bottom panels) to very low drag (top panels). {\bf b.} The diapycnal mixing associated with the diffusive terms in (\ref{eq:horizavgb}) decreases as the diffusivities are lowered. Here both the vertical and horizontal diffusivities are simultaneously decreased by factors of 2 and factors of 4, the lowest values (thick black curve) corresponding to run 3 (bottom panels in a).   \label{fig:profiles}}
\end{figure}

Having established that the 2LQG model has skill to predict the functional dependence of meridional transport in the Eady problem, we turn to the vertical structure of the buoyancy fluxes, a question that can be addressed only by going beyond 2LQG. An accurate description of the vertical structure of the transport tensor is crucial step in the development of a parameterization of baroclinic turbulence to be used in a fully 3D global ocean model.

\subsection{Depth-invariant meridional flux}

 
The QG framework readily provides a prediction for the vertical structure of the meridional buoyancy flux. According to equation (\ref{PVbulk}), the potential vorticity (\ref{QGPV}) is a material invariant of the undamped system. If the initial condition corresponds to vanishing PV, the PV will remain zero at subsequent times and the meridional PV flux vanishes at any depth. Within the QG framework this PV flux reads~\citep{Vallisbook}:
\begin{eqnarray}
0 = \overline{vq} = \overline{p_x p_{xx}} + \overline{p_x p_{yy}} + \frac{1}{\lambda^2} \overline{p_x p_{zz}} =  \frac{1}{\lambda^2} \partial_z (\overline{p_x p_z}) = \frac{1}{\lambda^2} \partial_z \overline{vb} \, ,
\end{eqnarray}
which indicates that the meridional buoyancy flux is depth-independent. 
\cor{This QG prediction turns out to be in very good agreement with the numerical data, as illustrated by the meridional buoyancy flux profiles in Figure~\ref{fig:profiles}: the low-drag profile in the top-left panel of Figure~\ref{fig:profiles}a is depth-independent to very good accuracy. The moderate-drag profile in the bottom-left panel of Figure~\ref{fig:profiles}a is also depth-independent in the interior, with slight departures restricted to a boundary-layer near the surface. Finally, solutions to the  QG system  -- equations (\ref{eq:rescaledQG3}-\ref{eq:rescaledQG4}) together with inversion relations (\ref{inversion1}-\ref{inversion2}) -- are characterized by depth-independent meridional flux profiles by construction.}

\subsection{Vertical flux and bulk stratification}


Having determined the vertical structure of the meridional buoyancy flux, we turn to the vertical buoyancy flux using relation (\ref{eq:horizavgb}). We have seen that the overall contribution from dissipative terms on the right-hand side of that relation is sometimes not entirely negligible, with ${r} \simeq 0.85$ instead of ${r}=1$, but how is that contribution distributed with depth? 

We answer that question in Figure~\ref{fig:profiles}, where we compare the two flux terms on the left-hand side of (\ref{eq:horizavgb}). Within the central half of the domain, these two terms balance each other to a good accuracy: the eddy transport current is directed along isopycnals in the bulk of the domain. However, for several runs this balance becomes unsatisfactory as we move closer to the top and bottom boundaries, where the vertical flux needs to be compatible with the boundary conditions: $w$ vanishes at the top free surface, while it matches the pumping velocity near the bottom boundary. In other words, the diffusive terms in (\ref{eq:horizavgb}) are not entirely negligible near the boundaries. In particular, we expect a strong contribution from the horizontal diffusive term in (\ref{eq:horizavgb}) at the surface, where surface quasi-geostrophic (SQG) dynamics predicts a forward cascade of buoyancy (towards small horizontal scales)~\citep{Held95,Lapeyre17}. The depth-independent meridional buoyancy flux in equation (\ref{eq:horizavgb}) is thus balanced by the vertical buoyancy flux term in the bulk of the domain and by the diffusive terms at the boundaries. The thickness of the surface-influenced region decreases with the diffusivities within the SQG framework. However, when the diffusivities reach very low values, sharp fronts develop, the latter being better described within the semi-geostrophic framework. Close inspection of our numerical runs indicates that the surface-influenced region is small only if the diffusivities are very weak, with large Reynolds (and P\'eclet) numbers, in a regime with weak isopycnal slopes and $Ro \ll 1$. Figure~\ref{fig:profiles} provides profiles from a numerical run in such a QG regime with extremely weak diffusivities. Both the meridional and the vertical buoyancy flux are depth-independent to a good approximation. The scaled buoyancy fluxes in figure~\ref{fig:profiles} are approximately equal in the bulk of the domain, which shows that the buoyancy flux is directed along isopycnals.
We also provide the profile of the mixing term $\epsilon_b=E_{b;z}\overline{(\partial_z b)^2} + E_{b;\perp} \overline{|\boldsymbol{\nabla}_{\perp} b|^2}$ for a series of runs where we lower both the vertical and horizontal diffusivities for otherwise constant parameters. We observe that there is significant diapycnal mixing near the top boundary, over a depth that decreases (albeit slowly) as the diffusivities are lowered. As the latter become asymptotically small, we expect the diapycnal mixing to be confined to the immediate vicinity of the upper boundary, the eddy-induced buoyancy current being then along isopycnals in most of the fluid domain.

The vertical structure of the emergent stratification is deduced from the horizontal- and time-average of the buoyancy equation (\ref{eq:eqbadim}):
\begin{eqnarray}
\frac{\mathrm{d} \overline{b}}{\mathrm{d}z} & = & \frac{\overline{wb} - Ro \int_0^z \overline{v}(\tilde{z}) \mathrm{d}\tilde{z}}{E_{b;z}} \simeq  \frac{\overline{wb} }{E_{b;z}}  \, ,
\end{eqnarray}
where we have neglected the very weak contribution from $\overline{v}$ (for the two runs displayed in Figure~\ref{fig:profiles}, this contribution is smaller than the contribution from $\overline{wb}$ by more than two orders of magnitude). The emergent stratification profile is thus readily given by that of the vertical buoyancy flux. In the bulk of the domain, the emergent stratification is thus uniform in the low-diffusivity strongly-QG regime where $\overline{wb}$ is depth-independent. 

\subsection{Gradual barotropization}

It is worth stressing the fact that the depth-invariance of the fluxes arises even when the turbulent kinetic energy profile (TKE) retains some significant depth dependence, as dictated by the QG result and illustrated by the profiles for a $\kappa_* ={\cal O}(1)$ run in the bottom panels in Figure~\ref{fig:profiles}a ($\kappa_*\simeq 0.6$). For lower drag, $\kappa_*={\cal O}(10^{-1})$, the flow barotropizes with a TKE profile that gradually becomes depth-invariant, as seen in the upper panels in Figure~\ref{fig:profiles}a ($\kappa_* \simeq 0.14$). The approach to the asymptotic low-drag predominantly barotropic state is thus rather slow, and this slow approach can be understood from QG vortex gas dynamics: multiply the uniform bulk QG PV (\ref{QGPV}) with $\tilde{b}=\partial_z \tilde{p}$, where the tilde denotes departure from the horizontal and temporal mean, before averaging horizontally and over time. After a few integrations by parts one is led to:
\begin{eqnarray}
\overline{{\bf v}^2}(z)-\frac{\overline{\tilde{b}^2}(z)}{\lambda^2} = \text{const.} \,  \label{relv}
\end{eqnarray}
The vortex gas scaling regime predicts that ${\bf v}^2$ is asymptotically greater than $\tilde{b}^2$ at low drag. The second term on the left-hand side of (\ref{relv}) is thus negligible at low drag and one concludes that the flow is barotropic. However, the approach to this asymptotically barotropic state is rather slow as $\kappa_*$ decreases. Indeed, the vortex gas scalings state that the (appropriately non-dimensionalized) buoyancy variance scales as $D_*$, while the horizontal kinetic energy scales as $D_* \log D_*$ (see~\citet{Gallet2020} for the derivation of these scaling estimates). While $D_*$ increases very rapidly with decreasing drag -- as $\exp(c_2/\kappa_*)$, see equation (\ref{eq:diffusivityVG}) -- the ratio of the first over the second term on the left-hand side of (\ref{relv}) increases only as $\log D_* \sim 1/\kappa_*$, hence the slow approach to the asymptotic barotropic state.

The vortex gas scaling theory performs surprisingly well even in the absence of complete barotropization. To some extent, this can be understood from 2LQG dynamics: the vortex gas theory is based on the idea that a barotropic vortex gas distorts the background buoyancy gradient, inducing buoyancy fluctuations at the inter-vortex scale. Within 2LQG, this regime corresponds to the large-scale-dynamics approximation given by equations (32) and (33) in~\citet{Thompson06} (see also \citet{Salmon80} and \citet{Larichev95}), where the buoyancy field is stirred by the barotropic flow only. An interesting point is that complete barotropization is not required for this large-scale-dynamics approximation to hold. Instead, only scale separation between the inter-vortex distance and the vortex core radius is necessary.

Of course, partial barotropization also affects the main power integral entering the theory, that is, the balance between the rate of release of available potential energy -- proportional to the dimensionless meridional buoyancy flux $D_*$ -- and the frictional dissipation term. Assuming that the latter corresponds mainly to dissipation by the barotropic flow, the rate of frictional dissipation scales as $\kappa_* D_* \log D_*$ (where the logarithmic correction stems from the fast velocities arising in the vicinity of the vortex cores, see~\citet{Gallet2020}). Balancing this estimate for the frictional dissipation rate with the rate $D_*$ of release of APE leads to the scaling prediction (\ref{eq:diffusivityVG}) for $D_*$. The assumption that frictional dissipation is associated with the barotropic flow is questionable when barotropization is only partial. In the  present 3D model, the frictional dissipation rate is given by the product of the effective friction coefficient with the squared bottom velocity. According to the relation (\ref{relv}), the bottom squared velocity may depart from the squared barotropic velocity by a correction proportional to $\overline{b^2}(0)$, the latter scaling as $D_*$ in the vortex gas theory. Including this correction leads to the frictional dissipation rate being proportional to $\kappa_* D_* [\log(D_*) + \text{const.} ] = \kappa_* D_* \log(D_*/\text{const.})$. Balancing this last expression for the frictional dissipation rate with the dimensionless rate of release of APE, $D_*$, leads again to the scaling prediction (\ref{eq:diffusivityVG}) for the meridional diffusivity. We conclude that partial barotropization does not affect the scaling prediction for the meridional buoyancy flux. 

In summary, it seems that the vortex gas theory only requires that the gas be dilute: the mixing length must be much greater than $\lambda$, the ratio of the two scaling as $\sqrt{D_*}$. Using the scaling expression (\ref{eq:diffusivityVG}) for $D_*$, we conclude that the vortex gas scaling predictions hold provided $\exp(\text{const.}/\kappa_*) \gg 1$, whereas strong barotropization of the flow requires the much more stringent criterion $\log D_* \sim 1/\kappa_* \gg 1$.

\section{Conclusion\label{sec:conclusion}}

Through a suite of numerical runs of the Boussinesq Eady problem, we have shown that the strongly stratified rapidly rotating regime is governed by quasi-geostrophic dynamics. One consequence is that the dependence of the transport properties of the turbulent flow on all control parameters is encoded in a scaling relation between the dimensionless diffusivity and the dimensionless drag. That scaling relation is correctly captured by the vortex gas scaling theory, see equation (\ref{eq:diffusivityVG}), initially put forward in the context of the 2LQG model and validated against numerical simulations with both linear and quadratic drag. In the context of the Eady problem, the two adjustable parameters arising in that scaling relation take slightly different values than in the 2LQG problem. We stress the fact that adjusting these two parameters leads to a quantitative theory for the magnitude and vertical structure of the meridional buoyancy flux, the vertical buoyancy flux and the emergent stratification, in terms of the background shear, the Coriolis parameter, the background stratification, the bottom friction and the diffusivities.

Indeed, the estimate (\ref{eq:diffusivityVG}) for the magnitude of the meridional buoyancy flux readily translates into a quantitative prediction for the vertical buoyancy flux through the standard along-isopycnal-transport argument. This argument is validated by comparing the numerical value of ${r}$ with the theoretical value ${r}=1$: we obtain numerically ${r} \simeq 0.85$ over the suite of simulations, with ${r} \to 1$ as we enter the very-low-diffusivity QG regime. The vertical buoyancy flux in turn leads to a quantitative prediction for the magnitude of the emergent stratification, set by a balance between eddy-induced vertical buoyancy transport and diffusion. This prediction for the emergent stratification allows us to determine the region of parameter space where the emergent stratification is negligible as compared to the imposed background stratification, and the region of parameter space where the sole emergent stratification is sufficient to induce QG dynamics in the absence of a background stratification.

Beyond scaling analysis, the 3D Eady model allows one to address the vertical structure of the eddy-induced transport. QG theory predicts that the meridional buoyancy flux is depth-invariant. This prediction is very specific to the Eady model, however, as it stems from the absence of interior PV gradients. One expects some vertical structure for the meridional buoyancy flux in a more realistic model with interior PV gradients. The prediction of a depth-invariant meridional flux is well satisfied by the numerical data (more so in the low-drag regime). The along-isopycnal-transport argument then implies that the vertical buoyancy flux is also depth-independent. The latter prediction is indeed validated by the strongly QG low-diffusivity data, although some depth-dependence remains within small boundary layers near the top and bottom boundaries (the thickness of which decreases, albeit rather slowly, with the diffusivities). A depth-independent vertical buoyancy flux in turn induces a uniform emergent stratification. The turbulent kinetic energy profile retains some depth dependence for moderate drag and gradually becomes more depth-invariant as the drag is decreased, in line with a barotropization of the flow. This slow barotropization of the flow as the drag coefficient decreases can be explained within the vortex gas scaling theory, and we have provided arguments indicating that some baroclinicity does not invalidate the theory. In particular, the theory could have skill to describe the transport induced by the mid-ocean eddies considered in \citet{Arbic2004b} based on current-meter data~(\citet{Wunsch1997} and references therein), which retain some significant baroclinicity. For instance, the bottom panels in Figure~\ref{fig:profiles}a illustrate an equilibrated flow with comparable baroclinic and barotropic turbulent kinetic energies. Such near equipartition of barotropic and baroclinic energies is consistent with ACC dynamics. The dimensionless drag coefficient is $\kappa_* \simeq 0.6$ for this run. Assuming the ocean-like parameter values $Ro=0.25$ and $N/f=30$, this value for $\kappa_*$ translates into a dimensionless linear drag coefficient $\kappa_{\text{eff}}=0.005$. Using a Coriolis parameter $f\simeq10^{-4}$~rad.s$^{-1}$ the corresponding dimensional magnitude of the linear drag coefficient is $(25 \, \text{days})^{-1}$. The order of magnitude is correct, this value being greater by a factor of four than the typical estimates reported in~\citet{Arbic2004b}. \cor{Vorticity snapshots associated with this run are provided in Figure~\ref{fig:snapshots}d,e. We observe that the vortex gas is moderately dilute, with an inter-vortex distance comparable to the core radius. The inter-vortex distance increases to 3 to 5 core radii for the lower-drag run illustrated in Figure~\ref{fig:snapshots}b,c. These values for the inter-vortex distance are within the range of typical oceanic observations.} Beyond the 2LQG approach of~\citet{Arbic2004b}, a more quantitative comparison with ACC dynamics and mid-ocean eddies in general would require \cor{the inclusion of} the vertical structure of the current profile and density stratification into the 3D model, together with quadratic bottom drag.

\begin{figure}
  \centerline{\includegraphics[width=0.6 \textwidth]{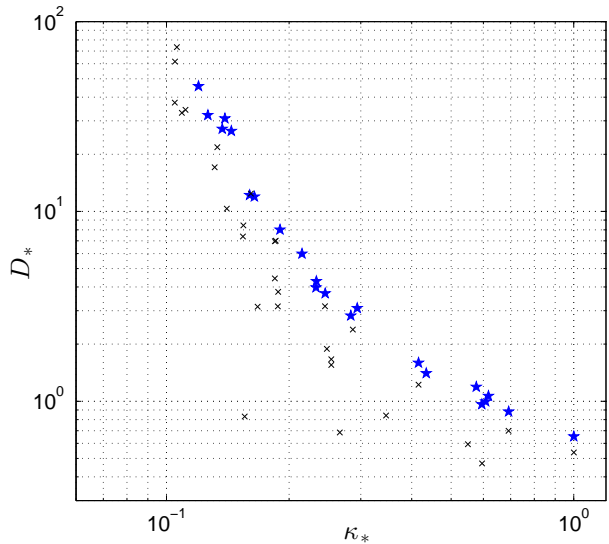}}
  \caption{Same as Figure~\ref{fig:Dstar_vs_kappastar} for the entire dataset. The blue stars denote the runs retained throughout the manuscript and reported in Table~\ref{Tableruns}. They are characterized by large enough Reynolds numbers, $Re_v > 4000$ and $Re_h>350$, together with low enough core vorticity to ensure QG dynamics, $\la vb \ra / (Ro \lambda^2) < 0.5$. The runs that violate any of these inequalities are represented with black crosses and reported in Table~\ref{Tableremainingruns}. They are characterized by a lower eddy diffusivity as compared to the QG vortex-gas master curve in Figure~\ref{fig:Dstar_vs_kappastar}.\label{fig:Dstar_vs_kappastar_extended}}
\end{figure}

Of course, whether the equilibrated Eady flow is governed by QG vortex gas dynamics depends on the values of the control parameters, and the runs in Table~\ref{Tableruns} represent only a subset of our entire suite of numerical runs. Satisfactory collapse onto the QG vortex-gas master curve in Figure~\ref{fig:Dstar_vs_kappastar} arises provided (i) the diffusivities are sufficiently low and (ii) the assumptions of QG dynamics are well satisfied. Specifically, for a run to be retained in Table~\ref{Tableruns} we demand that the Reynolds numbers be large enough, and that the isopycnal slope and vortex core vorticity be low enough. Regarding the low diffusivities, we used a threshold value for the vertical Reynolds (and P\'eclet) number $Re_v=Ro/E_z$ of the base shear flow, $Re_v > 4000$, together with a threshold value for the emergent horizontal Reynolds (and P\'eclet) number $Re_h=\sqrt{\la {\bf v}^2 \ra} \lambda / E_{\perp} > 350$. Regarding the assumptions of QG dynamics, we find that the isopycnal slope is small in all the numerical runs. However, the bulk vertical vorticity $\zeta$ reaches large values inside the vortex cores, especially in the low-drag regime\cor{, see Figure~\ref{fig:snapshots}b}. Within the vortex gas theory, the circulation of the vortices scales like the meridional eddy diffusivity $\la vb \ra / Ro$, while the vortex core radius scales like the Rossby deformation radius $\lambda$. Demanding that the core vorticity \cor{be at most comparable to} $f$ then amounts to satisfying the inequality $\la vb \ra / (Ro \lambda^2) < 0.5$, where the threshold value $0.5$ has been chosen somewhat arbitrarily. The runs in Table~\ref{Tableruns} satisfy the three inequalities above. The emergent Rossby number, measured as the rms vertical vorticity at mid-depth divided by $f$, is at most of the order of $0.1$ for these runs. The remaining runs are reported in Table~\ref{Tableremainingruns}. They violate at least one of the three inequalities above, which results in an eddy diffusivity that lies below the QG vortex-gas master curve, as illustrated in Figure~\ref{fig:Dstar_vs_kappastar_extended}. An extreme example of breakdown of the QG model when the assumptions of QG dynamics are not satisfied is reported in \citet{Molemaker}. They perform numerical simulations without bottom friction. In the absence of a large-scale energy sink for the QG inverse cascade, the flow intensifies until the emergent Rossby number is of the order of unity. The QG expansion then breaks down and, consistently, their Boussinesq simulations greatly depart from QG dynamics, with a forward energy cascade that halts the intensification of the flow.

In summary, our study demonstrates how the vortex-gas scaling theory can be extended to describe the structure and magnitude of eddy-induced transport by fully 3D baroclinic turbulence. The skill of the vortex-gas scaling theory for both the 2LQG model and the Eady model shows that the theoretical predictions hold regardless of whether the flow is driven by boundary dynamics or interior PV gradients. The model is strongly idealized, however, and should be complexified if the resulting parameterization is to describe real atmospheres or oceans. In the atmospheric context, one may want to include $\beta$ and address the eddy-induced transport in the equilibrated Charney model. The simple advective-diffusive balance that sets the emergent vertical stratification could be modified as well, with the goal of better capturing the feedback of baroclinic turbulence on the vertical structure of the atmosphere~\citep{HeldSuarez}.

$\beta$-plane dynamics induced by planetary curvature or sloping topography are also important ingredients of ocean dynamics. 
In the context of the ACC, an accurate description of mesoscale transport probably requires a base state with depth-dependent zonal shear and an inversion of the meridional potential vorticity gradient at some finite depth~\citep{Charney62,Pedloskybook,Smith09,Abernathey10}. The vortex gas approach has been extended to the $\beta$-plane within the 2LQG framework, but the skill of the resulting theory remains to be assessed for such a vertically-structured fully 3D model.

\subsubsection*{Acknowledgments}

This research is supported by the European Research Council under grant agreement FLAVE 757239. The numerical study was performed using HPC resources from GENCI-CINES and TGCC (grant 2021-A0102A12489 and grant 2021-A0102A10803).

\subsubsection*{Declaration of interests}

The authors report no conflict of interest.

\bibliographystyle{jfm}
\bibliography{Eady_bib_v2}

\appendix

\section{Effective friction coefficient associated with partial Ekman pumping\label{app:friction}}

The frictional boundary condition (\ref{bottomBC}) induces damping both directly, at the bottom boundary, and indirectly, because it leads to some vertical dependence of the flow and therefore some additional viscous dissipation. Indeed, a truncated Ekman spiral connects the bulk flow to the frictional bottom boundary condition. The additional viscous damping associated with the truncated Ekman spiral extracts energy from the bulk flow. In this appendix, we compute the resulting total friction acting on the bulk flow and the associated effective friction coefficient.

Consider a bulk horizontal flow $(U,V,0)$ connected to the bottom boundary condition (\ref{bottomBC}) by an Ekman spiral. In the vicinity of the bottom boundary the dimensionless boundary-layer equations are:
\begin{eqnarray}
\cor{V-v} & = & {E}_z \partial_{zz} u  \, , \\
\cor{u-U} & = & {E}_z \partial_{zz} v \, .
\end{eqnarray}
\cor{where the $z$-independent $U$ and $V$ terms arise from the horizontal pressure gradient.}
The solution that satisfies the bottom boundary condition is:
\begin{equation}
u+iv  =  (U+iV) F(z) \, ,
\end{equation}
where:
\begin{equation}
F(z) = 1- \frac{e^{-\frac{1+i}{\sqrt{2 {E}_z}}z}}{1+(1+i)\frac{\sqrt{{E}_z}}{\sqrt{2}\, \kappa}} \, . \label{Ffun}
\end{equation}
The effective dimensionless friction coefficient $\kappa_\text{eff}$ is defined as the total dissipated power associated with this flow divided by the kinetic energy of the bulk flow. It contains two terms, $\kappa_\text{eff} = \kappa_1 + \kappa_2$, where $\kappa_1$ is associated with the power directly dissipated by frictional damping at the bottom boundary while $\kappa_2$ is associated with viscous damping in the partial Ekman spiral (\ref{Ffun}). $\kappa_1$ is simply given by:
\begin{equation}
\kappa_1= \frac{{\kappa} |u(z=0)+i v(z=0)|^2}{|U+iV|^2} = {\kappa} |F(0)|^2 \, ,
\end{equation}
while $\kappa_2$ is given by:
\begin{equation}
\kappa_2= \frac{{E}_z \int_0^1 (\partial_z u)^2 + (\partial_z v)^2 \mathrm{d}z }{|U+iV|^2} \simeq {E}_z \int_0^\infty |F'(z)|^2 \mathrm{d}z  = \frac{\sqrt{{E}_z/2}}{\left( 1+ \sqrt{\frac{{E}_z}{2}}\frac{1}{\kappa} \right)^2+\frac{{E}_z}{2 \kappa^2}} \, .
\end{equation}

An alternative approach to determining $\kappa_\text{eff}$ consists in computing the friction term induced by Ekman pumping associated with a bulk flow that varies slowly with $x$ and $y$. Denoting the bulk vertical vorticity as $\zeta$, the pumping velocity just above the partial Ekman spiral reads  $w|_{0^+}=\kappa_3 \zeta|_{0^+}$. The goal here is to show that $\kappa_3$ is equal to $\kappa_\text{eff}$ above.

For simplicity, consider a purely barotropic flow $U(y)$ that varies slowly in $y$. The bulk flow has vertical vorticity $\zeta=-U'(y)$. The total flow, including the partial Ekman spiral, is:
\begin{equation}
u+iv  =  U(y)  F(z) \, ,
\end{equation}
and the incompressibility condition imposes a vertical velocity such that:
\begin{equation}
\partial_z w  =  -U'(y) \, \text{Im}\{F(z) \} \, .
\end{equation}
Integrating over the Ekman layer thickness yields the pumping velocity:
\begin{equation}
w|_{0^+}  =  \zeta|_{0^+}  \int_0^\infty \text{Im}\{F(z) \} \mathrm{d}z \, ,
\end{equation}
and we conclude that:
\begin{equation}
\kappa_3  =  \int_0^\infty \text{Im} \{F(z) \} \mathrm{d}z = \frac{\sqrt{2 {E}_z}}{1+ \frac{\sqrt{2 {E}_z}}{\kappa}+\frac{{E}_z}{\kappa^2}} \times \left(\frac{1}{2}+ \sqrt{\frac{{E}_z}{2}} \frac{1}{\kappa} \right) \, .
\end{equation}
One can check that the two approaches yield the same effective friction coefficient $\kappa_\text{eff}$, that is $\kappa_\text{eff}=\kappa_3=\kappa_1+\kappa_2$. 

\section{Quasi-geostrophy for the 3D Eady problem\label{app:QG}}

\subsection{Asymptotic expansion}
QG dynamics stems from an expansion in small isopycnal slope $Ro/\lambda^2 \ll 1$. To alleviate notations, we reproduce the expansion here by assuming a uniform stratification with $\lambda={\cal O}(1)$ and expanding the dimensionless variables in terms of the Rossby number $Ro \ll 1$ (but the reader should keep in mind that the results hold as long as $Ro/\lambda^2 \ll 1$). Expand the fields as:
\begin{eqnarray}
\tilde{{\bf v}} & = & Ro \, {\bf v}^{(0)} + Ro^2 \, {\bf v}^{(1)} + \dots \, , \\
\tilde{p} & = & Ro\, p^{(0)} +  Ro^2\, p^{(1)} + \dots \, , \\
\tilde{b} & = & Ro\, b^{(0)} +  Ro^2\, b^{(1)} + \dots \, ,
\end{eqnarray}
where the tilde on the left-hand side denote departures from the horizontally averaged fields. The fields depend on space and on the slow time variable $\hat{t}=Ro \, t$ only. We neglect the diffusive terms in the bulk of the fluid domain. The friction coefficient in the pumping boundary condition $w|_{0^+}=\kappa_\text{eff} \, \zeta|_{0^+}$ is assumed to be of order $Ro$, that is $\kappa_\text{eff}=Ro \, \hat{\kappa}$. Substitution into equation (\ref{eq:eqvadim}) yields, to order $Ro$:
\begin{eqnarray}
{\bf e}_z \times {\bf v}^{(0)} & = & - \bnabla p^{(0)} + b^{(0)} \, {\bf e}_z \label{eq:tempappB}
\end{eqnarray}
The boundary conditions reduce to $w^{(0)}|_{0^+}=w^{(0)}|_{1}=0$ at this order, and the solution corresponds to geostrophic and hydrostatic balances:
\begin{eqnarray}
{\bf v}^{(0)} = -\bnabla \times ( {p^{(0)} {\bf e}_z} ) \, , \qquad b^{(0)} = \partial_z p^{(0)} \, . 
\end{eqnarray}
The vertical velocity vanishes at this order\cor{, as seen by taking the vertical component of the curl of (\ref{eq:tempappB}) before making use of the incompressibility constraint and of the boundary conditions.} The vertical vorticity is $\zeta^{(0)}=\Delta_\perp p^{(0)}$.
Consider then the vertical vorticity equation and the buoyancy equation, both at order $Ro^2$:
\begin{eqnarray}
\partial_{\hat{t}} \Delta_\perp p^{(0)} + z \partial_x \Delta_\perp p^{(0)} + J(p^{(0)},\Delta_\perp p^{(0)}) & = & \partial_z w^{(1)} \, , \\
\partial_{\hat{t}} \partial_z p^{(0)} - \partial_x  p^{(0)} +z  \partial_{xz}  p^{(0)} + J(p^{(0)},\partial_z p^{(0)}) & = & -\lambda^2 w^{(1)} \label{eq:appborder1}
\end{eqnarray}
Differentiating the second equation with respect to $z$ before adding the first equation multiplied by $\lambda^2$ allows us to eliminate $w^{(1)}$. The resulting equation can be written as a conservation equation:
\begin{equation}
\partial_{\hat{t}} q + J(p^{(0)},q) +  \, z \, \partial_x q  =  0 \label{eq:appPVbulk}
\end{equation}
for the QG potential vorticity:
\begin{equation}
q = \bnabla_\perp^2 p^{(0)} +  \frac{\partial_{zz} p^{(0)}}{\lambda^2} \, . \label{eq:appPVdef}
\end{equation}
In terms of the original variables, equations (\ref{eq:appPVbulk}) and (\ref{eq:appPVdef}) correspond to (\ref{PVbulk}) and (\ref{QGPV}) in the main body of the article. The boundary conditions at this order are $w^{(1)}|_1=0$ and $w^{(1)}|_{0^+} = \hat{\kappa} \, \Delta_\perp p^{(0)}|_{0^+}$. Substitution of these values of $w^{(1)}$ into equation (\ref{eq:appborder1}) evaluated at $z=1$ and $z=0^+$ leads to the boundary conditions (\ref{BCQG1}) and (\ref{BCQG0}) in terms of the original variables.

\subsection{Inversion relation and effective 2D dynamics}

An efficient way to solve the QG problem is to march in time the 2D equations (\ref{eq:rescaledQG3}) and (\ref{eq:rescaledQG4}). To wit, at each time step we need to infer $P|_{0^+}$ and $P|_{1}$ in terms of $\partial_z P|_{0^+}$ and $\partial_z P|_{1}$. This inversion is performed by assuming that the PV vanishes in the bulk of the fluid domain, $Q=0$ being a trivial solution to (\ref{eq:rescaledQG1}). Decompose $P$ as a Fourier series in the horizontal directions:
\begin{eqnarray}
P(X,Y,z,T) = \Sigma_{\bf k} \hat{P}_{\bf k} (z,T) \times  e^{i {\bf k}\cdot {\bf X}} \, ,
\end{eqnarray}
where ${\bf X}=(X,Y)$ and the wave vector ${\bf k}$ takes the discrete values compatible with the periodic horizontal boundary conditions. We also introduce the following notations for the Fourier transform of $P$ and its vertical derivatives at the top and bottom boundaries:
\begin{eqnarray}
P|_{0^+} & = & \Sigma_{\bf k} \{P|_0\}_{\bf k} \, e^{i {\bf k}\cdot {\bf X}} \, , \\
P|_{1} & = & \Sigma_{\bf k} \{P|_1\}_{\bf k} \, e^{i {\bf k}\cdot {\bf X}} \, , \\
\partial_z P|_{0^+} & = & \Sigma_{\bf k} \{\partial_z P|_0\}_{\bf k} \, e^{i {\bf k}\cdot {\bf X}} \, , \\
\partial_z P|_{1} & = & \Sigma_{\bf k} \{\partial_z P|_1\}_{\bf k} \, e^{i {\bf k}\cdot {\bf X}} \, ,
\end{eqnarray}
where we omit the time dependence of the Fourier coefficients to alleviate notations. The PV associated with each wave vector ${\bf k}$ vanishes:
\begin{eqnarray}
-k^2 \hat{P}_{\bf k} + \partial_{zz} \hat{P}_{\bf k} = 0 \, ,
\end{eqnarray}
where $k=|{\bf k}|$. The solution to this equation whose vertical derivative matches $\{\partial_z P|_0\}_{\bf k}$ and $\{\partial_z P|_1\}_{\bf k}$ at the boundaries is:
\begin{eqnarray}
\hat{P}_{\bf k}(z,T) & = & \frac{- \{\partial_z P|_0\}_{\bf k}  \cosh[k(z-1)] + \{\partial_z P|_1\}_{\bf k}  \cosh (kz) }{ k \sinh k} \, ,
\end{eqnarray}
Evaluating this quantity at the boundaries $z=0$ and $z=1$ yields the two inversion relations providing $\{P|_0\}_{\bf k}$ and $\{P|_1\}_{\bf k}$ in terms of $\{\partial_z P|_0\}_{\bf k}$ and $\{\partial_z P|_1\}_{\bf k}$:
\begin{eqnarray}
\{P|_0\}_{\bf k} & = & \frac{ - \{\partial_z P|_0\}_{\bf k}  \cosh k + \{\partial_z P|_1\}_{\bf k} }{ k \sinh k} \, , \label{inversion1}\\
\{P|_1\}_{\bf k} & = & \frac{ - \{\partial_z P|_0\}_{\bf k} + \{\partial_z P|_1\}_{\bf k}  \cosh k }{ k \sinh k} \, . \label{inversion2}
\end{eqnarray}
A time step of the QG Eady problem consists in (i) marching the 2D equations (\ref{eq:rescaledQG3}) and (\ref{eq:rescaledQG4}) for one timestep, before (ii) inferring the new fields $P|_{0^+}$ and $P|_1$ using the inversion relations (\ref{inversion1}) and (\ref{inversion2}).

\subsection{Symmetry argument for $\overline{wu}$\label{app:sym}}

Consider a solution $\tilde{p}={\cal P}(x,y,z,t)$, $q={\cal Q}(x,y,z,t)$ to the QG system: the functions ${\cal P}$ and ${\cal Q}$ satisfy equations (\ref{PVbulk}-\ref{QGPV}) with the boundary conditions (\ref{BCQG1}) and (\ref{BCQG0}). The vertical velocity associated with this solution is inferred from equation (\ref{eq:appborder1}), which reads in terms of the original variables:
\begin{eqnarray}
-\lambda^2 w & = & \partial_{t z}  \tilde{p} - Ro \, \partial_x  \tilde{p} +Ro\, z  \partial_{xz}  \tilde{p} + J(\tilde{p},\partial_z \tilde{p}) \, . \label{diagnosticw}
\end{eqnarray}
Introduce the following transformed solution:
\begin{eqnarray}
\hat{p}(x,\hat{y},z,t)=-{\cal P}(x,y=-\hat{y},z,t)\, , \qquad \hat{q}(x,\hat{y},z,t)=-{\cal Q}(x,y=-\hat{y},z,t) \, . \label{transformedsol}
\end{eqnarray}
The fields $\hat{p}(x,\hat{y},z,t)$ and $\hat{q}(x,\hat{y},z,t)$ also satisfy the QG system. To check this, the operator $\partial_y$ should be understood as a derivative with respect to the second variable, and thus replaced by $\partial_{\hat{y}}$ in equations (\ref{PVbulk}-\ref{QGPV}) and boundary conditions (\ref{BCQG1}) and (\ref{BCQG0}) where $\tilde{p}$ and $q$ have been replaced by $\hat{p}$ and $\hat{q}$, respectively. There is a direct correspondence between the fields associated with the original solution and the fields associated with the transformed one (the latter being evaluated at $\hat{y}=-y$). Denoting the former with the standard notations and the latter with hat variables, we obtain:
\begin{eqnarray}
\hat{u} & = & -\partial_{\hat{y}} \hat{p} = - \partial_y {\cal P} = u \, , \\
\hat{v} & = & \partial_{x} \hat{p} = - \partial_x {\cal P} = -v \, , \\
\hat{b} & = & \partial_{z} \hat{p} = - \partial_z {\cal P} = -b \, , \\
\hat{w} & = & -w \, ,
\end{eqnarray}
where the hat variables are evaluated at $\hat{y}=-y$, while the variables associated with the original solution are evaluated at $y$. The last relation has been obtained by substitution into equation (\ref{diagnosticw}) for the vertical velocity.
This correspondence indicates, for instance, that the meridional and vertical buoyancy fluxes are equal for the original and for the transformed solutions: $\overline{\hat{v} \hat{b}}=\overline{vb}$ and $\overline{\hat{w} \hat{b}}=\overline{wb}$. However, the vertical flux of zonal momentum is opposite between the two solutions: $\overline{\hat{w} \hat{u}}=-\overline{wu}$. Because the original solution and the transformed ones are equally probable, we conclude that $\overline{wu}=0$ within the QG framework, while $\overline{vb} \neq 0$ and $\overline{wb} \neq 0$.

\begin{table}
\hspace{-1.9 cm}
\begin{tabular}{c|c|c|c|c|c|c|c|c|c|c|c|c}
{Run} & {$Ro$} & {$(N/f)^2$} & {$L/H$} & {$E_z$} & {$E_{\perp}$} & {$\lambda$} & {$\kappa$} & {$\kappa_*$} & {$D_*$} & {$\lambda \left<wb \right>/Ro^3$} & {$\left<{\bf v}^2\right>$} & {$n_x \plh n_y \plh n_z$} \\ \hline
1 & 0.025 & 40000 & 4000 & 2.50$\plh10^{-6}$ & 0.01 & 200.00 & 0.00008 & 0.617 & 1.066 & 1.006 & 0.00236 & $256\plh256\plh128$ \\ 
2 & 0.025 & 40000 & 4000 & 1.25$\plh10^{-6}$ & 0.005 & 200.00 & 0.00008 & 0.608 & 0.997 & 0.929 & 0.00263 & $256\plh256\plh128$\\ 
3 & 0.025 & 40000 & 4000 & 6.25$\plh10^{-7}$ & 0.0025 & 200.00 & 0.00008 & 0.595 & 0.964 & 0.931 & 0.00293 & $512\plh512\plh128$\\ 
4 & 0.25 & 1000 & 2000 & 1.00$\plh10^{-5}$ & 0.15 & 42.45 & 0.0011 & 0.144 & 26.519 & 21.888 & 7.739 & $512\plh 512\plh 128$\\ 
5 & 0.25 & 1000 & 1000 & 1.00$\plh10^{-5}$ & 0.04 & 33.54 & 0.02 & 0.294 & 3.096 & 2.767 & 0.793 & $256\plh 256\plh 256$\\ 
6 & 0.5 & 10000 & 2500 & 1.00$\plh10^{-4}$ & 0.15 & 100.04 & 0.01 & 1.000 & 0.652 & 0.600 & 0.657 & $256\plh 256\plh 48$\\ 
7 & 0.5 & 10000 & 2500 & 1.00$\plh10^{-4}$ & 0.15 & 100.05 & 0.005 & 0.692 & 0.882 & 0.816 & 0.777 & $256\plh 256\plh 48$\\ 
8 & 0.5 & 10000 & 2500 & 1.00$\plh10^{-4}$ & 0.15 & 100.09 & 0.0025 & 0.416 & 1.594 & 1.486 & 1.323 & $256\plh 256\plh 48$\\ 
9 & 2.5 & 0 & 4000 & 1.00$\plh10^{-4}$ & 5 & 115.68 & 0.005 & 0.160 & 12.191 & 10.015 & 301.108 & $512\plh 512\plh 92$\\ 
10 & 2.5 & 0 & 2500 & 2.00$\plh10^{-4}$ & 2 & 79.21 & 0.01 & 0.190 & 8.013 & 6.221 & 183.354 & $256\plh 256\plh 128$\\ 
11 & 2.5 & 10000 & 5000 & 5.00$\plh10^{-4}$ & 1.2 & 105.15 & 0.007 & 0.233 & 4.291 & 3.738 & 91.618 & $512\plh 512\plh 24$\\ 
12 & 2.5 & 10000 & 2500 & 5.00$\plh10^{-4}$ & 1.2 & 101.92 & 0.02 & 0.434 & 1.405 & 1.267 & 27.619 & $512\plh 512\plh 24$\\ 
13 & 2.5 & 10000 & 2500 & 1.00$\plh10^{-4}$ & 2.4 & 127.50 & 0.007 & 0.215 & 5.983 & 5.081 & 146.612 & $256\plh 256\plh 48$\\ 
14 & 2.5 & 10000 & 2500 & 1.00$\plh10^{-4}$ & 2.4 & 114.96 & 0.02 & 0.284 & 2.823 & 2.364 & 61.876 & $256\plh 256\plh 48$\\ 
15 & 2.5 & 10000 & 2500 & 5.00$\plh10^{-4}$ & 2.4 & 104.97 & 0.007 & 0.233 & 3.973 & 3.406 & 81.832 & $256\plh 256\plh 48$\\ 
16 & 2.5 & 30000 & 5000 & 5.00$\plh10^{-4}$ & 1.2 & 174.96 & 0.004 & 0.245 & 3.699 & 3.426 & 78.460 & $512\plh 512\plh 24$\\ 
17 & 2.5 & 30000 & 10000 & 5.00$\plh10^{-4}$ & 5 & 186.09 & 0.0018 & 0.126 & 32.131 & 27.615 & 970.795 & $1024\plh 1024\plh 48$\\ 
18 & 2.5 & 30000 & 10000 & 1.00$\plh10^{-4}$ & 5 & 237.11 & 0.0014 & 0.120 & 45.663 & 39.816 & 1656.548 & $512\plh 512\plh 48$\\ 
19 & 2.5 & 30000 & 10000 & 1.00$\plh10^{-4}$ & 5 & 217.34 & 0.0018 & 0.137 & 27.226 & 23.464 & 879.013 & $256\plh 256\plh 48$\\ 
20 & 2.5 & 30000 & 10000 & 5.00$\plh10^{-4}$ & 5 & 178.39 & 0.0025 & 0.164 & 11.956 & 10.431 & 284.298 & $1024\plh 1024\plh 48$\\ 
21 & 2.5 & 30000 & 10000 & 1.00$\plh10^{-4}$ & 5 & 220.64 & 0.0018 & 0.139 & 30.874 & 26.574 & 961.958 & $1024\plh 1024\plh 96$\\
\end{tabular}

\caption{Control and emergent parameter values for the runs retained in the main study. For all the runs $E_{b;z}=E_z$ and $E_{b;\perp}=E_{\perp}$. The last column indicates the number of modes in each direction used for the numerical simulation.\label{Tableruns}}
\end{table}

\begin{table}
\hspace{-1.1 cm}
\begin{tabular}{c|c|c|c|c|c|c|c|c|c|c|c|c}
{Run} & {$Ro$} & {$(N/f)^2$} & {$L/H$} & {$E_z$} & {$E_{\perp}$} & {$\lambda$} & {$\kappa$} & {$\kappa_*$} & {$D_*$} & {$\lambda \left<wb \right>/Ro^3$} & {$\left<{\bf v}^2\right>$} & {$n_x \plh n_y \plh n_z$} \\ \hline
22 & 2.5 & 5000 & 2500 & 0.0025 & 1.2 & 71.05 & 0.03 & 0.550 & 0.59 & 0.54 & 8.07 & $512\plh512\plh24$\\ 
23 & 2.5 & 5000 & 2500 & 0.0025 & 1.2 & 71.75 & 0.01 & 0.248 & 1.89 & 1.69 & 29.26 & $512\plh512\plh24$\\ 
24 & 2.5 & 10000 & 2500 & 0.0025 & 1.2 & 100.25 & 0.01 & 0.346 & 0.84 & 0.80 & 10.27 & $512\plh512\plh24$\\ 
25 & 2.5 & 10000 & 2500 & 0.0025 & 1.2 & 101.08 & 0.005 & 0.188 & 3.77 & 3.49 & 65.25 & $512\plh512\plh24$\\ 
26 & 2.5 & 10000 & 2500 & 0.0025 & 1.2 & 100.14 & 0.02 & 0.596 & 0.47 & 0.45 & 5.77 & $512\plh512\plh24$\\ 
27 & 2.5 & 10000 & 5000 & 0.0025 & 1.2 & 108.67 & 0.0025 & 0.105 & 37.45 & 31.34 & 1105.12 & $512\plh512\plh24$\\ 
28 & 2.5 & 10000 & 2500 & 0.0025 & 1.2 & 100.46 & 0.007 & 0.254 & 1.55 & 1.46 & 21.49 & $512\plh512\plh24$\\ 
29 & 2.5 & 10000 & 2500 & 0.0025 & 1.2 & 102.04 & 0.004 & 0.154 & 7.38 & 6.71 & 154.24 & $512\plh512\plh24$\\ 
30 & 2.5 & 10000 & 5000 & 0.0025 & 1.2 & 102.28 & 0.004 & 0.154 & 8.44 & 7.54 & 173.93 & $512\plh512\plh24$\\ 
31 & 2.5 & 30000 & 5000 & 0.0025 & 1.2 & 173.53 & 0.0025 & 0.167 & 3.15 & 3.08 & 46.31 & $512\plh512\plh24$\\ 
32 & 2.5 & 10000 & 5000 & 0.0025 & 1.2 & 104.31 & 0.0033 & 0.131 & 17.09 & 14.74 & 405.84 & $512\plh512\plh24$\\ 
33 & 2.5 & 10000 & 2500 & 0.0025 & 1.2 & 100.89 & 0.005 & 0.188 & 3.16 & 2.90 & 53.70 & $512\plh512\plh24$\\ 
34 & 2.5 & 5000 & 2500 & 0.0025 & 1.2 & 73.01 & 0.007 & 0.185 & 4.44 & 3.85 & 80.99 & $512\plh512\plh24$\\ 
35 & 2.5 & 5000 & 2500 & 0.0025 & 1.2 & 75.63 & 0.005 & 0.141 & 10.34 & 8.66 & 237.14 & $512\plh512\plh24$\\ 
36 & 2.5 & 10000 & 5000 & 0.01 & 1.2 & 100.05 & 0.007 & 0.266 & 0.68 & 0.68 & 3.01 & $512\plh512\plh24$\\ 
37 & 0.25 & 10000 & 2500 & 0.00025 & 0.12 & 100.00 & 0.007 & 2.018 & 1.27 & 1.23 & 0.17 & $512\plh512\plh24$\\ 
38 & 2.5 & 10000 & 5000 & 0.01 & 1.2 & 100.06 & 0.004 & 0.156 & 0.83 & 0.81 & 4.55 & $512\plh512\plh24$\\ 
39 & 2.5 & 10000 & 2500 & 0.0025 & 2.4 & 100.48 & 0.007 & 0.254 & 1.66 & 1.54 & 21.47 & $512\plh512\plh24$\\ 
40 & 2.5 & 30000 & 10000 & 0.0005 & 5 & 196.06 & 0.0014 & 0.105 & 61.57 & 52.96 & 2252.80 & $256\plh256\plh48$\\ 
41 & 0.5 & 10000 & 2500 & 0.0001 & 1.2 & 100.02 & 0.01 & 1.000 & 0.54 & 0.38 & 0.41 & $256\plh256\plh48$\\ 
42 & 0.5 & 10000 & 2500 & 0.0001 & 1.2 & 100.03 & 0.005 & 0.692 & 0.70 & 0.52 & 0.46 & $256\plh256\plh48$\\ 
43 & 0.5 & 10000 & 2500 & 0.0001 & 1.2 & 100.06 & 0.0025 & 0.416 & 1.22 & 0.95 & 0.71 & $256\plh256\plh48$\\ 
44 & 2.5 & 30000 & 10000 & 0.0005 & 5 & 174.64 & 0.004 & 0.245 & 3.17 & 2.78 & 57.48 & $256\plh256\plh48$\\ 
45 & 0.25 & 1000 & 1000 & $10^{-5}$ & 0.08 & 32.72 & 0.02 & 0.287 & 2.39 & 2.06 & 0.57 & $256\plh256\plh256$\\ 
46 & 0.25 & 1000 & 1000 & $10^{-5}$ & 0.3 & 36.57 & 0.0016 & 0.161 & 12.43 & 9.36 & 2.74 & $256\plh256\plh256$\\ 
47 & 2.5 & 0 & 10000 & 0.0001 & 5 & 209.83 & 0.0014 & 0.106 & 73.30 & 60.96 & 2779.47 & $256\plh256\plh48$\\ 
48 & 2.5 & 0 & 4000 & 0.0001 & 5 & 92.71 & 0.01 & 0.185 & 7.01 & 5.04 & 138.78 & $256\plh256\plh92$\\ 
49 & 2.5 & 0 & 4000 & 0.0001 & 5 & 92.30 & 0.01 & 0.185 & 6.96 & 5.02 & 139.72 & $256\plh256\plh92$\\ 
50 & 2.5 & 0 & 4000 & 0.0001 & 5 & 161.36 & 0.002 & 0.111 & 34.28 & 28.29 & 1263.37 & $256\plh256\plh92$\\ 
51 & 2.5 & 0 & 4000 & 0.0001 & 5 & 158.00 & 0.002 & 0.109 & 33.07 & 26.13 & 1125.54 & $256\plh256\plh92$\\ 
52 & 0.25 & 1000 & 2000 & $10^{-5}$ & 0.6 & 39.21 & 0.0011 & 0.133 & 21.79 & 14.76 & 4.28 & $256\plh256\plh256$\\ 
\end{tabular}
\caption{Control and emergent parameter values for the remaining runs, characterized by either too large diffusivities (too low Reynolds number) or too large core vorticity as compared to $f$.\label{Tableremainingruns}}

\end{table}

\end{document}